\documentclass[journal,10pt]{IEEEtran}

\usepackage[T1]{fontenc}

\usepackage{cite}

\ifCLASSINFOpdf
   \usepackage[pdftex]{graphicx}
\else
\fi
\usepackage{graphicx}

\usepackage{amsmath}
\usepackage[cmintegrals]{newtxmath}
\newtheorem{lemma}{Lemma}

\usepackage{blkarray}

\begin{document}
%
\title{Signal-Aligned Network Coding \\ in K-User MIMO Interference Channels \\ with Limited Receiver Cooperation}
%
%
%

\author{Tse-Tin~Chan,~\IEEEmembership{Student~Member,~IEEE,}
        and~Tat-Ming~Lok,~\IEEEmembership{Senior~Member,~IEEE}
\thanks{This work was supported in part by the
General Research Fund from the Research Grants Council of the Hong Kong SAR under Project CUHK 14203616. This work has been submitted to the {IEEE} for possible publication. Copyright may be transferred without notice, after which this version may no longer be accessible. The material in this paper was presented  in part at the {IEEE/CIC} International Conference on Communications in China, Beijing, China, 2018~\cite{Signal-Aligned Network Coding in Interference Channels with Limited Receiver Cooperation}.}
\thanks{The authors are with the Department of Information Engineering, The Chinese University of Hong Kong, Hong Kong (e-mail: ctt014@ie.cuhk.edu.hk; tmlok@ie.cuhk.edu.hk).}}

\maketitle

\begin{abstract}
In this paper, we propose a signal-aligned network coding (SNC) scheme for K-user time-varying multiple-input multiple-output (MIMO) interference channels with limited receiver cooperation. We assume that the receivers are connected to a central processor via wired cooperation links with individual limited capacities. Our SNC scheme determines the precoding matrices of the transmitters so that the transmitted signals are aligned at each receiver. The aligned signals are then decoded into noiseless integer combinations of messages, also known as network-coded messages, by physical-layer network coding. The key idea of our scheme is to ensure that independent integer combinations of messages can be decoded at the receivers. Hence the central processor can recover the original messages of the transmitters by solving the linearly independent equations. We prove that our SNC scheme achieves full degrees of freedom (DoF) by utilizing signal alignment and physical-layer network coding. Simulation results show that our SNC scheme outperforms the compute-and-forward scheme in the finite SNR regime of the two-user and the three-user cases. The performance improvement of our SNC scheme mainly comes from efficient utilization of the signal subspaces for conveying independent linear equations of messages to the central processor.
\end{abstract}


%
\IEEEpeerreviewmaketitle

\section{Introduction}
\label{Introduction}
\IEEEPARstart{D}{ue} to the broadcast nature of wireless medium, the signals heard by a receiver are not only the signals from the intended transmitter and the noise, but also the interfering signals from other nearby transmitters. With the rapid growth of mobile device usage, interference has become a bottleneck in today's wireless networks. In order to improve the efficiency of wireless communications, coordinated multipoint (CoMP) has been proposed for Long Term Evolution Advanced (LTE-Advanced) networks. For uplink transmissions, joint reception and processing is one of the CoMP techniques in which the receivers share the received analog signals or the raw received signal samples among themselves via cooperation links to perform joint decoding. However, in some practical situations, the cooperation links do not support the high requirements of analog transmission or raw signal samples forwarding due to the limited capacities of the cooperation links. Therefore, other advanced schemes for handling the signals heard by the receivers are desired.

In this paper, we focus on interference channels with limited receiver cooperation. We assume the receivers are connected to a central processor via independent wired cooperation links with individual restricted capacities. A special case of our channel model is the scenario in which the receivers are interconnected through cooperation links and one of the receivers acts as the central processor. As forwarding analog signal samples from the receivers to the central processor generates excessive overhead, we focus on the situation that the cooperation links forward digital decoded packets instead of analog signal samples. The traffic of the limited receiver cooperation remains comparable to the wireless throughput. The interference channel with limited receiver cooperation has been widely investigated in many researches such as those about cloud radio access network (C-RAN), distributed multiple-input multiple-output (MIMO) system, wireless local area network (WLAN), etc. This architecture is generally viewed as a promising candidate to improve the network performance by efficient utilization of the wired cooperation links~\cite{Multi-cell MIMO cooperative networks: A new look at interference}.

\subsection{Literature Review}
Physical-layer network coding (PNC)~\cite{Hot topic: Physical-layer network coding} brings the promising idea of network coding (NC)~\cite{Network information flow, Linear network coding} from network layer to physical layer to ease the interference problem in wireless channels. PNC was originally proposed for two-way relay channels (TWRC) in which two users exchange information with the aid of an intermediate relay. PNC demodulates superimposed signals into network-coded data by utilizing the additive property of electromagnetic (EM) waves. PNC makes use of the network-coded data and users' own transmitted data for exchanging data streams between users. 

Under channel models similar to ours, there are many approaches proposed which make use of the wireless interference for network coding. We describe two of them and introduce some related papers in this subsection.

\subsubsection{Compress-and-forward}
Compress-and-forward transforms the channel into a virtual multiple access channel (MAC). The receivers compress the received signals via Wyner-Ziv coding or other compression strategies. The central processor decompresses the compressed signals forwarded from the receivers and recovers the original messages of the transmitters. Nevertheless, the network performance is deteriorated by the quantization noise. Also, some compress-and-forward schemes are difficult to be implemented because of the high complexity.

Sanderovich \textit{et al.}~\cite{Uplink macro diversity of limited backhaul cellular network} derived the achievable rates for an uplink cellular network with joint multicell processing. The cell-site was only interfered by the users of the adjacent cells rather than by all users of the cells cooperated. Park \textit{et al.}~\cite{Joint decompression and decoding for cloud radio access networks} studied uplink multi-antenna C-RANs and proposed a joint decompression and decoding scheme. The cloud decoder jointly decompressed the compressed signals forwarded from the base stations and decoded the signals of the mobile stations. The result gives us an important insight into understanding the rate which can be achieved by compress-and-forward. The high complexity of joint decompression and decoding is a hindrance to practical implementation. Zhou and Yu~\cite{Uplink multicell processing with limited backhaul via per-base-station successive interference cancellation} proposed a scheme with lower complexity compared with joint decoding for an uplink multicell joint processing model. The scheme performed Wyner-Ziv compress-and-forward on a per-base-station basis followed by successive interference cancellation (SIC) at the central processor. Zhou and Yu~\cite{Optimized backhaul compression for uplink cloud radio access network} studied an uplink C-RAN and focused on optimizing the quantization noise levels at all base stations for weighted sum-rate maximization under a sum backhaul capacity constraint. They proposed an algorithm for allocating the backhaul capacities. This sum backhaul capacity constraint is particularly suited to wireless backhauls which are implemented based on an orthogonal access scheme.

\subsubsection{Compute-and-forward}
In compute-and-forward, the receivers decode the superimposed signals into noiseless linear equations, in a finite field, of the transmitted messages. The integer coefficients in the equations, also known as network coding coefficients, are close to the corresponding channel coefficients. After collecting sufficient linearly independent equations forwarded from the receivers, the central processor can recover the original messages of the transmitters. However, there are two major problems impairing the performance of compute-and-forward. First, the rank deficiency occurs if the equations forwarded from the receivers are linearly dependent. As a result, the central processor is unable to recover the original messages. Second, there exists a rate penalty which comes from the non-integer parts of the channel coefficients during decoding. The received signals are scaled so that the scaled received linear combination of codewords is close to an integer linear combination. However, this scaling factor also results in amplification of the noise which decreases the achievable rates.

Nazer and Gastpar~\cite{Compute-and-forward: Harnessing interference through structured codes} proposed compute-and-forward and its example in a two-user distributed MIMO network. Wei and Chen~\cite{Compute-and-forward network coding design over multi-source multi-relay channels} considered maximizing the transmission rate by designing the network coding coefficients in a multi-source multi-relay system. Hong and Caire~\cite{Compute-and-forward strategies for cooperative distributed antenna systems} considered a distributed antenna system with equal-capacity backhaul links. For the uplink system, they proposed a scheme which selected a subset of the receivers so that the matrix formed by the integer combinations of the transmitted messages had full rank and the sum rate was maximized. Soussi, Zaidi, and Vandendorpe~\cite{Compute-and-forward on a multi-user multi-relay channel} proposed an algorithm for a multi-user multi-relay channel. They focused on designing the precoding factor of the transmitters so as to increase the full rank probability of the matrix formed by the linear equations of the transmitted messages and maximize the transmission rate. Yang \textit{et al.}~\cite{A linear network coding approach for uplink distributed MIMO systems: Protocol and outage behavior} proposed an approach aimed at reducing the outage probability by designing the network coding coefficients for an uplink distributed MIMO system. Chen, Fan, and Letaief~\cite{Compute-and-forward: Optimization over multisource-multirelay networks} studied the maximization problem of  the multicast throughput by proper resource allocations for given coefficient vectors. In general, none of compress-and-forward and compute-and-forward outperforms another one in all regimes. 

Some researches such as~\cite{Degrees of freedom of the MIMO Y channel: signal space alignment for network coding, A signal-space aligned network coding approach to distributed MIMO, Signal alignment: Enabling physical layer network coding for MIMO networking, Interference alignment with physical-layer network coding in MIMO relay channels} showed that PNC can be employed together with signal alignment to achieve better performance. Signal alignment was proposed in~\cite{Degrees of freedom of the MIMO Y channel: signal space alignment for network coding} based on the idea of interference alignment (IA)~\cite{Degrees of freedom region of the MIMO X channel, Communication over MIMO X channels, Interference alignment and degrees of freedom of the K-user interference channel} to investigate the degrees of freedom (DoF) of the MIMO Y channel, in which three users exchanged messages with each other via a relay. Yang, Yuan, and Sun~\cite{A signal-space aligned network coding approach to distributed MIMO} studied an uplink distributed MIMO system with a sum backhaul rate constraint. They proposed a scheme making use of signal alignment and PNC for the block-fading channel, in which the channel coefficients remained constant over a block of symbols. Moreover, the performance of their scheme is limited by the numbers of antenna per node.

Furthermore, \cite{Aligned interference neutralization and the degrees of freedom of} and \cite{Degrees of freedom of two-hop wireless networks: Everyone gets the entire cake} focus on utilizing the advantages of signal alignment to tackle the interference problem in two-hop interference channels. However, SNC is a new transmission strategy in a different situation. Apart from the difference in the channel models, \cite{Aligned interference neutralization and the degrees of freedom of} and \cite{Degrees of freedom of two-hop wireless networks: Everyone gets the entire cake} consider the relays apply linear transformations to the analog signals so that the interfering analog signals are neutralized at the destinations. However, we consider that the receivers and the central processor are communicated through wired cooperation links. Only digital decoded packets can be forwarded via the wired limited cooperation links as forwarding analog signal samples would cause excessive traffic.

\subsection{Main Contributions}
We propose a signal-aligned network coding (SNC) scheme for $K$-user time-varying MIMO interference channels with limited receiver cooperation. The receivers are connected to a central processor through cooperation links with individual limited capacities. As the DoF, also known as the capacity pre-log or the multiplexing gain, provides a first-order approximation to the capacity, it plays an important role in characterizing the capacity behavior in the high signal-to-noise ratio (SNR) regime. We show that our proposed SNC scheme can achieve full DoF. In other words, at high SNR, our SNC scheme is able to achieve approximately the capacity of the channel in which unlimited cooperation among the receivers is allowed. This full DoF achievement is not limited by the number of antennas of each node. Furthermore, simulation results show that our SNC scheme outperforms the compute-and-forward scheme in the finite SNR regime of the two-user and the three-user cases.

Our SNC scheme utilizes physical-layer network coding and signal alignment. We consider that the receivers decode noiseless integer combinations of messages, also known as network-coded messages, from the transmitters. The decoding of noiseless linear equations of messages can be achieved by compute-and-forward or other PNC strategies. The received analog signals or the raw received signal samples cannot be forwarded from the receivers to the central processor due to the restricted capacities of the cooperation links. The capacity of each cooperation link is just sufficient to forward the decoded messages or the integer combinations of them in the same finite field. The signal alignment technique used in this paper is based on the precoding over multiple symbol extensions of the time-varying channel. Our SNC scheme designs the precoding matrices so that the interfering signals from the transmitters are aligned at each receiver and independent linear equations of the transmitted messages can be decoded at each receiver. As a result, the central processor can recover the original messages of the transmitters by solving the linearly independent equations. The performance improvement of our SNC scheme mainly comes from efficient utilization of the signal subspaces for conveying independent integer combinations of messages to the central processor. Moreover, our research shows that signal alignment is a useful technique to deal with the rank deficiency and the non-integer penalty problems in compute-and-forward schemes.

The rest of this paper is organized as follows. Section~\ref{System Model} describes the system model. Section~\ref{Illustrative Example} gives a two-user illustrative example of our SNC scheme. Section~\ref{SNC in K-User Time-Varying Interference Channels with Limited Receiver Cooperation} provides the generalized result of our scheme. Section~\ref{Simulation Results} shows the numerical results to evaluate our proposed scheme by comparing it with the compute-and-forward scheme. Section~\ref{Conclusion} concludes this paper. Finally, the detailed proofs of various results are given in Appendices~{\ref{SNC scheme for K users}--\ref{Degrees of Freedom for MIMO General Cases}}.

\subsection{Notations}
In this paper, letters of bold upper case, bold lower case, and lower case indicate matrices, vectors, and scalars respectively. $\mathbb{C}^{m \times n}$ denotes the set of all complex-valued $m \times n$ matrices. $\mathbb{F}^{m \times n}_q$ represents the set of all $m \times n$ matrices in a finite field of size~$q$. $\oplus$ and $\otimes$ mean the addition operation and multiplication operation, respectively, over that finite field. $\mathbb{Z}^{+}$ denotes the set of all positive integers and $\mathbf{I}_{d}$ indicates the $d \times d$ identity matrix. $\text{diag}(\cdot)$ represents a square diagonal matrix with the elements described inside the bracket. Moreover, $(\cdot)^{\text{T}}$, $(\cdot)^{\text{H}}$, $\|\cdot\|$, and $\mathbb{E}[\cdot]$ denote transpose, conjugate transpose, norm, and statistical expectation respectively.

\section{System Model}
\label{System Model}
\begin{figure}[!t]
 \centering
  \includegraphics[width=3.5in]{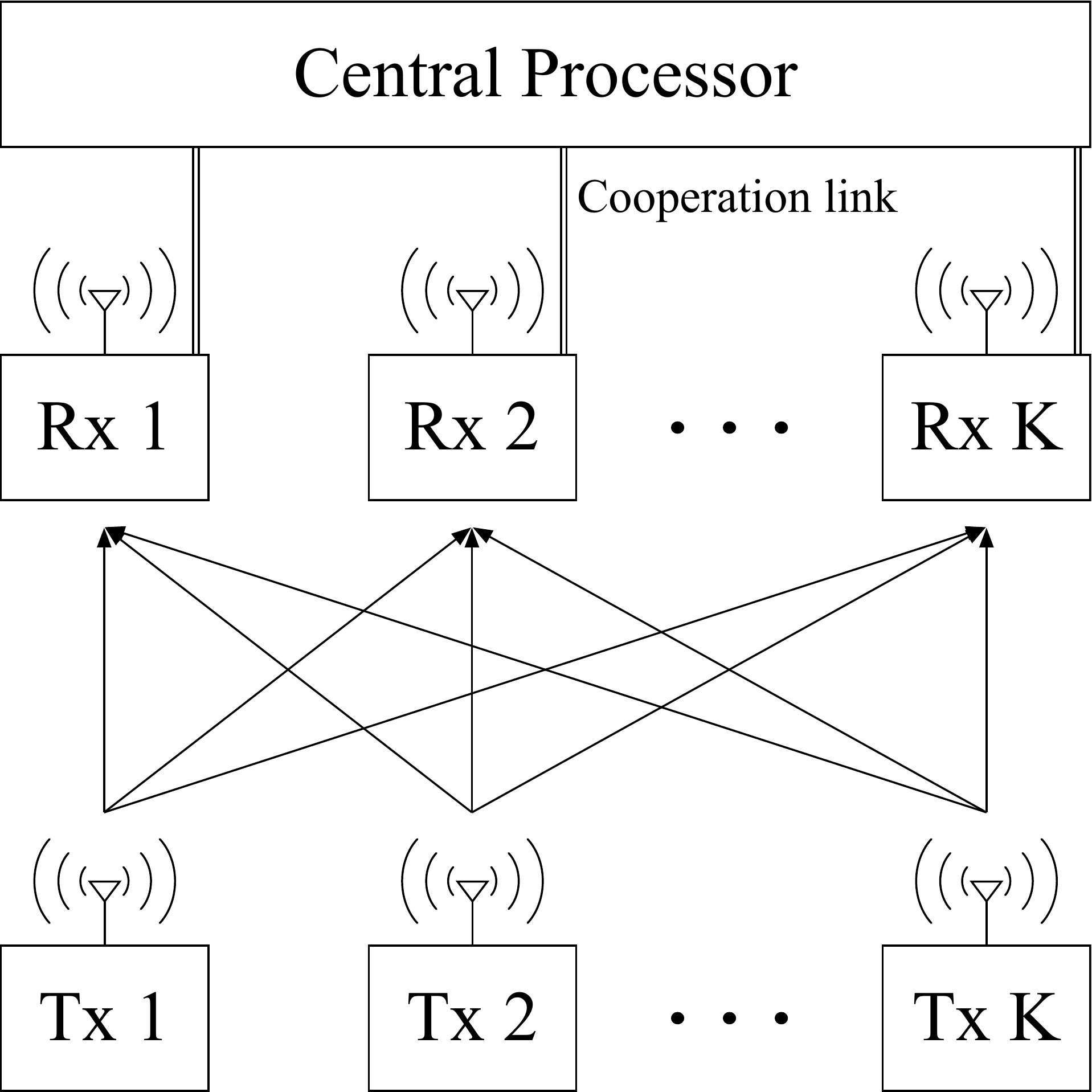}
  \caption{A $K$-user time-varying interference channel with limited receiver cooperation, which consists of $K$~transmitters (Tx), $K$~receivers (Rx), and a central processor.}  
  \label{A $K$-user interference channel with limited receiver cooperation.}
\end{figure} 
We consider a time-varying interference channel with limited receiver cooperation which consists of $K$~transmitters, $K$~receivers, and a central processor as illustrated in \figurename~\ref{A $K$-user interference channel with limited receiver cooperation.}. Each receiver is connected to the central processor via an independent noiseless cooperation link. Each transmitter and receiver has one antenna and nodes with multiple antennas are considered later in this paper. We assign unique indices ${k \in \{1,2,\dots,K\}}$ and ${l \in \{1,2,\dots,K\}}$ to each transmitter and receiver respectively. The overall transmission consists of two stages. In the first stage, the transmitters convey messages, which can be coded symbols, to the receivers in the interference channel. We assume that the transmitters send signals synchronously and share the same time, frequency, and code resources. We consider the interference from concurrent transmissions is much stronger than the noise. In the second stage, the receivers process and forward the received messages or the linear equations of them in the same finite field to the central processor via cooperation links. Having collected the messages from the receivers, the central processor recovers all original messages of the transmitters. In this paper, we assume that the rate-constraint of a cooperation link is comparable to the capacities of the links from each transmitter to the receiver connected to that cooperation link. In other words, the capacity of each cooperation link is just sufficient to forward the decoded messages rather than the raw received signals. Furthermore, we assume instantaneous channel state information (CSI) is globally available.

\subsection{Transmitters}
The system adopts an $N$ symbol extension of the channel where $N \in \mathbb{Z}^+$. The $N$ symbol extension means that the $N$ symbols transmitted from each transmitter over $N$ slots are collectively denoted as a supersymbol. In the first transmission stage, transmitter~$k$ modulates message vector ${\mathbf{b}_k(t) \in \mathbb{F}^{N \times 1}_q}$ to signal vector ${\mathbf{x}_k(t) \in \mathbb{C}^{N \times 1}}$ where
\begin{IEEEeqnarray}{rCCCl}
\label{b_k(t)}
\mathbf{b}_k(t) =
  \begin{bmatrix}
b_k^{(1)}(t)\\
b_k^{(2)}(t)\\
\vdots\\
b_k^{(N)}(t)
  \end{bmatrix} =
  \begin{bmatrix}
b_k(N(t-1)+1)\\
b_k(N(t-1)+2)\\
\vdots\\
b_k(Nt)
  \end{bmatrix}
\end{IEEEeqnarray}
and
\begin{IEEEeqnarray}{rCCCl}
\label{x_k(t)}
\mathbf{x}_k(t)  =
  \begin{bmatrix}
x_k^{(1)}(t)\\
x_k^{(2)}(t)\\
\vdots\\
x_k^{(N)}(t)
  \end{bmatrix} =
  \begin{bmatrix}
x_k(N(t-1)+1)\\
x_k(N(t-1)+2)\\
\vdots\\
x_k(Nt)
  \end{bmatrix}.
\end{IEEEeqnarray} The superscript~$(i)$ is used to denote the $i$-th slot of the $N$ symbol extension. Message vector $\mathbf{b}_{k}(t)$ is the $N$ symbol extension of independent and identically distributed (i.i.d.) message $b_{k}(t)$ introduced in (\ref{b_k(t)}). Signal vector $\mathbf{x}_{k}(t)$ is the $N$ symbol extension of signal $x_{k}(t)$ presented in (\ref{x_k(t)}) and $\mathbb{E}[\mathbf{x}_{k}(t)\mathbf{x}_{k}^{\text{H}}(t)] = \mathbf{I}_{N}$. In this paper, the time index~${t \in \mathbb{Z}^{+}}$ can be used to describe time slots, frequency slots, or time-frequency slots.

Transmitter~$k$ sends signal vector $\mathbf{x}_k(t)$ with linear precoding matrix ${\mathbf{V}_k(t) \in \mathbb{C}^{N \times N}}$ where
\begin{IEEEeqnarray}{rCl}
& & \mathbf{V}_k(t) \nonumber \\ & = &
  \begin{bmatrix}
\mathbf{v}_k^{(1)}(t) & \mathbf{v}_k^{(2)}(t) & \cdots & \mathbf{v}_k^{(N)}(t)
  \end{bmatrix} \nonumber \\ & = &
  \begin{bmatrix}
\mathbf{v}_k(N(t-1)+1) & \mathbf{v}_k(N(t-1)+2) & \cdots & \mathbf{v}_k(Nt)
  \end{bmatrix}.
\end{IEEEeqnarray} The $i$-th column vector of $\mathbf{V}_k(t)$, ${\mathbf{v}_k^{(i)}(t) \in \mathbb{C}^{N \times 1}}$, is the precoding vector for signal $x_k^{(i)}(t)$ presented in (\ref{x_k(t)}). Let $p_k^{(i)}(t)$ and $p_{k,\text{max}}^{(i)}(t)$ be the actual transmit power and the maximum transmit power for transmitting signal $x_k^{(i)}(t)$ respectively. The transmit power constraint is
\begin{equation}
0 \le p_k^{(i)}(t) = \|{\mathbf{v}_k^{(i)}(t)}\|^2 \le p_{k,\text{max}}^{(i)}(t).
\end{equation}

\subsection{Receivers}
The received signal of a receiver is the superposition of the signals from the transmitters weighted by their corresponding channels and the received signal is affected by the noise. With the $N$ symbol extension of the channel, receiver~$l$ observes received signal vector ${\mathbf{y}_l(t) \in \mathbb{C}^{N \times 1}}$ where
\begin{IEEEeqnarray}{rCl}
\label{{y}_l(t)}
\mathbf{y}_l(t) & = & 
  \begin{bmatrix}
y_l^{(1)}(t)\\
y_l^{(2)}(t)\\
\vdots\\
y_l^{(N)}(t)
  \end{bmatrix} =
  \begin{bmatrix}
y_l(N(t-1)+1)\\
y_l(N(t-1)+2)\\
\vdots\\
y_l(Nt)
  \end{bmatrix}
\nonumber \\ & = & \sum\limits_{k=1}^{K} \mathbf{H}_{l,k}(t) \mathbf{V}_k(t) \mathbf{x}_k(t) + \mathbf{n}_l(t).
\end{IEEEeqnarray}
Received signal vector $\mathbf{y}_l(t)$ is the $N$ symbol extension of received signal $y_l(t)$ presented in~(\ref{{y}_l(t)}). Diagonal channel matrix ${\mathbf{H}_{l,k}(t) \in \mathbb{C}^{N \times N}}$ is the $N$ symbol extension of channel coefficient $h_{l,k}(t)$ where
\begin{align}
\label{extended_channel}
&\mathbf{H}_{l,k}(t) \nonumber \\ = &
  \begin{bmatrix}
{h}_{l,k}^{(1)}(t) & 0 & \ldots & 0 \\
0 & {h}_{l,k}^{(2)}(t) & \ldots & 0 \\
\vdots & \vdots & \ddots & \vdots \\
0 & 0& \ldots & {h}_{l,k}^{(N)}(t)
  \end{bmatrix} \nonumber \\ = &
  \scriptstyle{
  \begin{bmatrix}
{h}_{l,k}(N(t-1)+1) & 0 & \ldots & 0 \\
0 & {h}_{l,k}(N(t-1)+2) & \ldots & 0 \\
\vdots & \vdots & \ddots & \vdots \\
0 & 0& \ldots & {h}_{l,k}(Nt)
  \end{bmatrix}}
\end{align}
and $h_{l,k}(t)$ denotes the CSI of the link from transmitter~$k$ to receiver~$l$. Moreover, noise vector ${\mathbf{n}_l(t) \in \mathbb{C}^{N \times 1}}$ is the $N$ symbol extension of noise term $n_l(t)$ with variance $\sigma^2_l(t)$ at receiver~$l$ where
\begin{IEEEeqnarray}{rCCCl}
\mathbf{n}_l(t) & = & 
  \begin{bmatrix}
n_l^{(1)}(t)\\
n_l^{(2)}(t)\\
\vdots\\
n_l^{(N)}(t)
  \end{bmatrix} =
  \begin{bmatrix}
n_l(N(t-1)+1)\\
n_l(N(t-1)+2)\\
\vdots\\
n_l(Nt)
  \end{bmatrix}.
\end{IEEEeqnarray} In this paper, we assume all channel coefficients are i.i.d. zero-mean unit-variance complex Gaussian random variables. Hence, channel matrix $\mathbf{H}_{l,k}(t)$ has rank~$N$ almost surely. We also assume all noise terms are i.i.d. complex additive white Gaussian noise (AWGN).

Receiver~$l$ applies linear filtering matrix ${\mathbf{U}_l(t) \in \mathbb{C}^{N \times N}}$ to received signal vector $\mathbf{y}_l(t)$ where
\begin{IEEEeqnarray}{rCl}
& & \mathbf{U}_l(t)  \nonumber \\ & = &
  \begin{bmatrix}
\mathbf{u}_l^{(1)}(t) & \mathbf{u}_l^{(2)}(t) & \cdots & \mathbf{u}_l^{(N)}(t)
  \end{bmatrix} \nonumber \\ & = &
  \begin{bmatrix}
\mathbf{u}_l(N(t-1)+1) & \mathbf{u}_l(N(t-1)+2) & \cdots & \mathbf{u}_l(Nt)
  \end{bmatrix}.
\end{IEEEeqnarray} The filtered signal vector of receiver~$l$ is ${\mathbf{x}'_l(t) \in \mathbb{C}^{N \times 1}}$ where
\begin{IEEEeqnarray}{rCl}
\mathbf{x}'_l(t) & = &
  \begin{bmatrix}
{x'_l}^{(1)}(t)\\
{x'_l}^{(2)}(t)\\
\vdots\\
{x'_l}^{(N)}(t)
  \end{bmatrix} =
  \begin{bmatrix}
{x'_l}(N(t-1)+1)\\
{x'_l}(N(t-1)+2)\\
\vdots\\
{x'_l}(Nt)
  \end{bmatrix} \nonumber \\
& = & \mathbf{U}_l^{\text{H}}(t) \mathbf{y}_l(t) \nonumber \\
& = & \sum\limits_{k=1}^{K} \mathbf{U}_l^{\text{H}}(t) \mathbf{H}_{l,k}(t) \mathbf{V}_k(t) \mathbf{x}_k(t) + \mathbf{U}_l^{\text{H}}(t) \mathbf{n}_l(t).
\end{IEEEeqnarray} Then receiver~$l$ demodulates filtered signal vector $\mathbf{x}'_l(t)$ to demodulated message vector ${\mathbf{b}'_l(t) \in \mathbb{F}^{N \times 1}_q}$ where
\begin{equation}
\mathbf{b}'_l(t) =
  \begin{bmatrix}
{b'_l}^{(1)}(t)\\
{b'_l}^{(2)}(t)\\
\vdots\\
{b'_l}^{(N)}(t)
  \end{bmatrix} = 
  \begin{bmatrix}
{b'_l}(N(t-1)+1)\\
{b'_l}(N(t-1)+2)\\
\vdots\\
{b'_l}(Nt)
  \end{bmatrix}.
\end{equation}

\subsection{Central Processor}
In the second transmission stage, receiver~$l$ forwards demodulated message vector $\mathbf{b}'_l(t)$ to the central processor through an independent noiseless cooperation link. In order to quantify the restriction on the cooperation links, we consider the capacities of each cooperation link are $\log_2(1+\bar{\rho})$ where $\bar{\rho}$ is the average SNR of the links in the interference channel, i.e., the average SNR of the links from each transmitter to each receiver. 

The central processor collects all forwarded messages from the receivers and then recovers all original messages of the transmitters. The recovered message vector of transmitter~$k$ is ${\mathbf{\hat{b}}_k(t) \in \mathbb{F}^{N \times 1}_q}$ where
\begin{IEEEeqnarray}{rCCCl}
\mathbf{\hat{b}}_k(t) =
  \begin{bmatrix}
\hat{b}_k^{(1)}(t)\\
\hat{b}_k^{(2)}(t)\\
\vdots\\
\hat{b}_k^{(N)}(t)
  \end{bmatrix} =
  \begin{bmatrix}
\hat{b}_k(N(t-1)+1)\\
\hat{b}_k(N(t-1)+2)\\
\vdots\\
\hat{b}_k(Nt)
  \end{bmatrix}.
\end{IEEEeqnarray} If $\mathbf{\hat{b}}_k(t) \neq \mathbf{b}_k(t)$ for any $k$, a decoding error occurs. For the sake of simplicity, the time index~$t$ is omitted in the rest of this paper.

The achievable rate of sending signal $x_{k}^{(n)}$ from transmitter $k$ to receiver $l$ is
\begin{equation}
{R}_{l,k}^{(n)} = \text{log}_2\left(1+{{(
\mathbf{u}_{l,k}^{(n)})^{\text{H}}(\mathbf{H}_{l,k})(\mathbf{v}_{l,k}^{(n)})(\mathbf{v}_{l,k}^{(n)})^{\text{H}}(\mathbf{H}_{l,k})^{\text{H}}(\mathbf{u}_{l,k}^{(n)})}\over{(\mathbf{u}_{l,k}^{(n)})^{\text{H}}({\sigma}_{k}^2\mathbf{I}_N)(\mathbf{u}_{l,k}^{(n)})}}\right),
\end{equation} where $\mathbf{v}_{l,k}^{(n)}$ and $\mathbf{u}_{l,k}^{(n)}$ are the precoding and decoding vectors, respectively, for transmitting {signal $x_{k}^{(n)}$} from transmitter $k$ to receiver $l$. The end-to-end sum-rate of the SNC scheme is described at the end of Section~\ref{SNC in K-User Time-Varying Interference Channels with Limited Receiver Cooperation}.

\subsection{Degrees of Freedom}
The capacity of transmitter~$k$ at SNR~$\rho$ can be expressed as
\begin{equation}
	C_k(\rho) = d_{k} \log_2(\rho)+\text{o}(\log_2(\rho)).
\end{equation}
$\text{o}(\log_2(\rho))$ is a function that $\frac{\text{o}(\log_2(\rho))}{\log_2(\rho)}$ tends to zero when $\rho$ tends to infinity. The capacity pre-log factor~$d_{k}$ is the DoF of transmitter~$k$, which is also known as the multiplexing gain. The DoF of transmitter~$k$ can be found by
\begin{IEEEeqnarray}{rCl}
d_{k} & = &  \lim_{\rho\to\infty} \frac{C_k(\rho)}{\log_2(\rho)}.
\end{IEEEeqnarray}

\section{Illustrative Example}
\label{Illustrative Example}
We present a simple example of our SNC scheme in this section. The detailed proof is shown in Section~\ref{SNC in K-User Time-Varying Interference Channels with Limited Receiver Cooperation}. We consider there are two transmitters and two receivers, i.e.,  ${K = L = 2}$. We show the users can achieve total $d_{\sum} = 5$ DoF over an ${N = 3}$ symbol extension in time-varying interference channels with limited receiver cooperation by our SNC scheme.	

First, transmitter~$1$ modulates message vector
\begin{equation}
  \mathbf{b}_1 =
  \begin{bmatrix}
b_1^{(1)}\\
b_1^{(2)}\\
b_1^{(3)}
  \end{bmatrix}
\end{equation}
to signal vector 
\begin{equation}
  \mathbf{x}_1 =
  \begin{bmatrix}
x_1^{(1)}\\
x_1^{(2)}\\
x_1^{(3)}
  \end{bmatrix}
\end{equation}
while transmitter~$2$ modulates message vector
\begin{equation}
  \mathbf{b}_2 =
  \begin{bmatrix}
b_2^{(1)}\\
b_2^{(2)}
  \end{bmatrix}
\end{equation}
to signal vector
\begin{equation}
  \mathbf{x}_2 =
  \begin{bmatrix}
x_2^{(1)}\\
x_2^{(2)}
  \end{bmatrix}.
\end{equation}
Afterward transmitter~${k \in \{1,2\}}$ sends signal vector $\mathbf{x}_{k}$ with linear precoding matrix $\mathbf{V}_{k}$.

The ${3 \times 1}$ received signal vector at receiver~${l \in \{1,2\}}$ is
\begin{equation}
\mathbf{y}_l =  \mathbf{H}_{l,1} \mathbf{V}_1 \mathbf{x}_1 +  \mathbf{H}_{l,2} \mathbf{V}_2 \mathbf{x}_2 + \mathbf{n}_l,
\end{equation}
where
\begin{equation}
\mathbf{H}_{l,k} =
  \begin{bmatrix}
{h}_{l,k}^{(1)} & 0 & 0 \\
0 & {h}_{l,k}^{(2)} & 0 \\
0 & 0 & {h}_{l,k}^{(3)}
  \end{bmatrix}
\end{equation}
and
\begin{equation}
  \mathbf{n}_l =
  \begin{bmatrix}
n_l^{(1)}\\
n_l^{(2)}\\
n_l^{(3)}
  \end{bmatrix}.
\end{equation}
In order to achieve the ideas of our SNC scheme, we can set the linear precoding matrices of the transmitters as
\begin{IEEEeqnarray}{rCl}
\mathbf{V}_1 & = & 
\begin{bmatrix}
\mathbf{G}_{1,2}^2\mathbf{w} & \mathbf{G}_{1,2}\mathbf{G}_{2,2}\mathbf{w} & \mathbf{G}_{2,2}^2\mathbf{w}
\end{bmatrix}, \\
\mathbf{V}_2 & = &
\begin{bmatrix}
\mathbf{G}_{1,2}\mathbf{w} & \mathbf{G}_{2,2}\mathbf{w}
\end{bmatrix}
\end{IEEEeqnarray}
where $\mathbf{G}_{1,2} = \mathbf{H}_{1,1}^{-1}\mathbf{H}_{1,2}$, $\mathbf{G}_{2,2} = \mathbf{H}_{2,1}^{-1}\mathbf{H}_{2,2}$, and 
\begin{equation}
  \mathbf{w} =
  \begin{bmatrix}
1\\
1\\
1
  \end{bmatrix}.
\end{equation}
As a result, $\mathbf{V}_1$ is a ${3 \times 3}$ matrix and $\mathbf{V}_2$ is a ${3 \times 2}$ matrix.

\begin{figure}[!t]
\centering
  \includegraphics[width=3.5in]{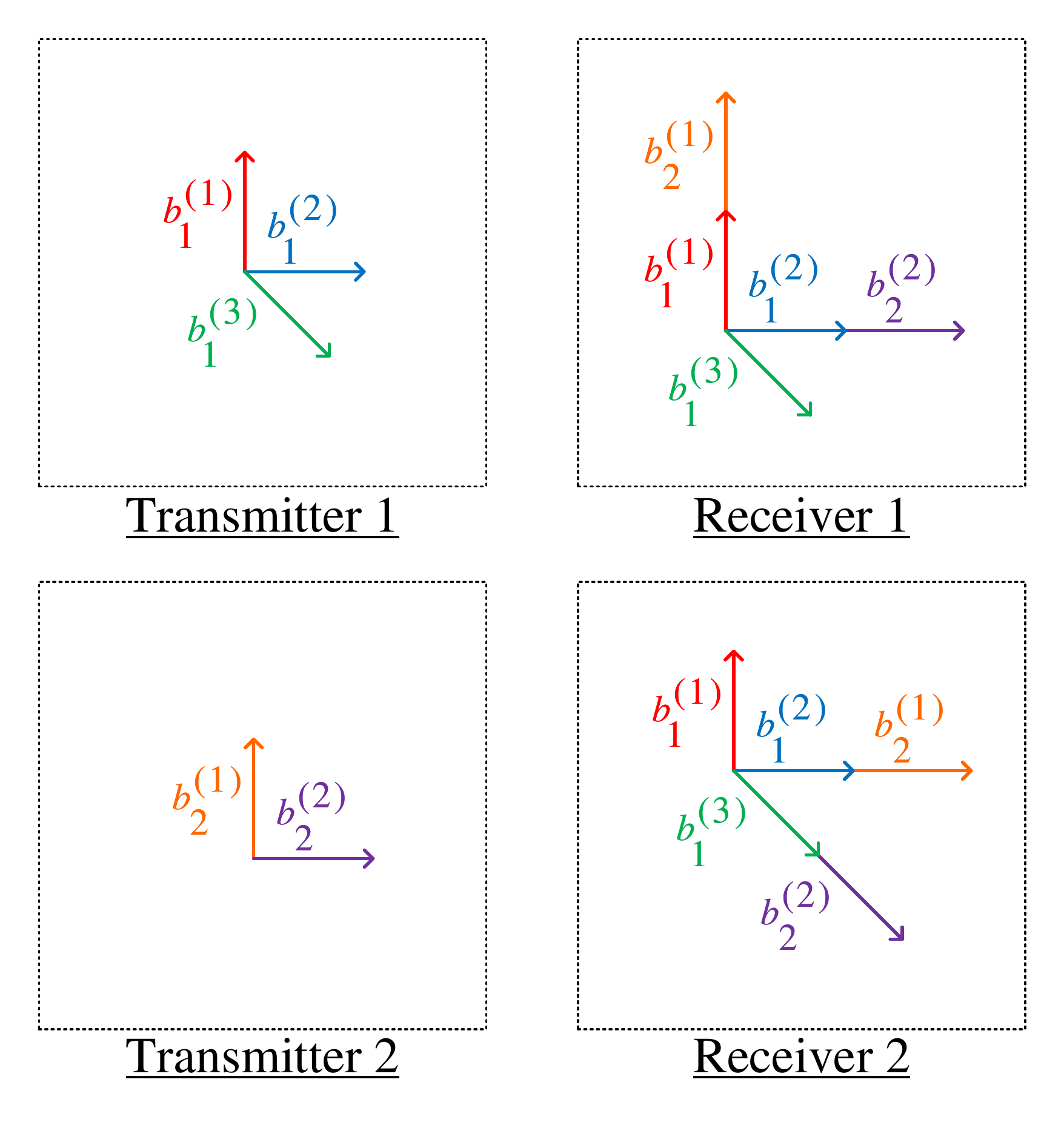}
  \caption{A 3-dimensional vector diagram illustrating the signal alignment of SNC in the two-user interference channel with limited receiver cooperation.}  
  \label{A 3-dimensional vector diagram illustrating the signal alignment of SNC in the two-user interference channel with limited receiver cooperation.}
\end{figure}

The messages decoded at the receivers are shown in \figurename~\ref{A 3-dimensional vector diagram illustrating the signal alignment of SNC in the two-user interference channel with limited receiver cooperation.}.
Now we describe how the linear precoding matrices of the transmitters affect the messages decoded at the receivers in details. As the multiplications of diagonal matrices are commutative, the multiplications of the channel matrices and the linear precoding matrices at receiver~$1$ are
\begin{align}
&\mathbf{H}_{1,1} \mathbf{V}_1 =
\begin{bmatrix}
\mathbf{H}_{1,2}\mathbf{G}_{1,2}\mathbf{w} & \mathbf{H}_{1,2} \mathbf{G}_{2,2} \mathbf{w} & \mathbf{H}_{1,1} \mathbf{G}_{2,2}^2 \mathbf{w}
\end{bmatrix}, \\ 
&\mathbf{H}_{1,2} \mathbf{V}_2 = 
\begin{bmatrix}
\mathbf{H}_{1,2} \mathbf{G}_{1,2} \mathbf{w} & \mathbf{H}_{1,2} \mathbf{G}_{2,2}\mathbf{w}
\end{bmatrix},
\end{align}
where $\mathbf{H}_{1,1}\mathbf{V}_1$ is a ${3 \times 3}$ matrix and $\mathbf{H}_{1,2} \mathbf{V}_2$ is a ${3 \times 2}$ matrix. The first column vector of $\mathbf{H}_{1,1} \mathbf{V}_1$, which is $\mathbf{H}_{1,2}\mathbf{G}_{1,2}\mathbf{w}$, for signal $x_1^{(1)}$ is the same as the first column vector of $\mathbf{H}_{1,2} \mathbf{V}_2$ for signal $x_2^{(1)}$. A similar relationship holds between signals $x_1^{(2)}$ and $x_2^{(2)}$ at receiver~$1$. Receiver~$1$ decodes received signal vector $\mathbf{y}_1$ by linear filtering matrix $\mathbf{U}_1^\text{H} = (\mathbf{H}_{1,1} \mathbf{V}_1)^{-1}$. The ${3 \times 1}$ filtered signal vector $\mathbf{x}'_1$ would be
\begin{IEEEeqnarray}{rCCCl}
\mathbf{x}'_1  & = & \mathbf{U}_1^{\text{H}} \mathbf{y}_1 & = & \mathbf{U}_1^{\text{H}} \mathbf{H}_{1,1} \mathbf{V}_1 \mathbf{x}_1 + \mathbf{U}_1^{\text{H}} \mathbf{H}_{1,2} \mathbf{V}_2 \mathbf{x}_2 + \mathbf{U}_1^{\text{H}} \mathbf{n}_1.
\end{IEEEeqnarray}

As the first two column vectors of $\mathbf{H}_{1,1} \mathbf{V}_1$ are the same as the column vectors of $\mathbf{H}_{1,2} \mathbf{V}_2$, we can express filtered signal vector $\mathbf{x}'_1$ as
\begin{equation}
\mathbf{x}'_1 =
  \begin{bmatrix}
x_1^{(1)} + x_2^{(1)}\\
x_1^{(2)} + x_2^{(2)}\\
x_1^{(3)}
  \end{bmatrix}
+ \mathbf{U}_1^{\text{H}} \mathbf{n}_1.
\end{equation}
We treat the aligned signal (e.g. ${x_1^{(1)} + x_2^{(1)}}$) as an unknown for demodulation rather than demodulate the original signals (e.g. $x_1^{(1)}$ and $x_2^{(1)}$) individually. This idea of PNC demodulation~\cite{Hot topic: Physical-layer network coding} also applies in the rest of this paper. Receiver~$1$ demodulates filtered signal vectors $\mathbf{x}'_1$ to network-coded message vector
\begin{equation}
\mathbf{b}'_1 =
  \begin{bmatrix}
b_1^{(1)} \oplus b_2^{(1)}\\
b_1^{(2)} \oplus b_2^{(2)}\\
b_1^{(3)}
  \end{bmatrix}.
\end{equation}
Signal $x_1^{(3)}$ is demodulated by conventional demodulation while the other signals are demodulated by PNC demodulation.

Afterward we look at the signals filtered and demodulated at receiver~$2$.  As the multiplications of diagonal matrices are commutative, the multiplications of the channel matrices and the linear precoding matrices at receiver~$2$ are
\begin{align}
&\mathbf{H}_{2,1} \mathbf{V}_1 =
\begin{bmatrix}
\mathbf{H}_{2,1} \mathbf{G}_{1,2}^2 \mathbf{w} & \mathbf{H}_{2,2} \mathbf{G}_{1,2} \mathbf{w} & \mathbf{H}_{2,2} \mathbf{G}_{2,2} \mathbf{w}
\end{bmatrix}, \\
&\mathbf{H}_{2,2} \mathbf{V}_2 = 
\begin{bmatrix} 
\mathbf{H}_{2,2} \mathbf{G}_{1,2} \mathbf{w} & \mathbf{H}_{2,2} \mathbf{G}_{2,2} \mathbf{w}
\end{bmatrix},
\end{align}
where $\mathbf{H}_{2,1} \mathbf{V}_1$ is a ${3 \times 3}$ matrix and $\mathbf{H}_{2,2} \mathbf{V}_2$ is a ${3 \times 2}$ matrix. The last two column vectors of $\mathbf{H}_{2,1} \mathbf{V}_1$ for signals $x_1^{(2)}$ and $x_1^{(3)}$ are the same as the column vectors of $\mathbf{H}_{2,2} \mathbf{V}_2$ for signals $x_2^{(1)}$ and $x_2^{(2)}$. Receiver $2$ decodes received signal vector $\mathbf{y}_2$ through linear filtering matrix $\mathbf{U}_2^\text{H} = (\mathbf{H}_{2,1} \mathbf{V}_1)^{-1}$. The ${3 \times 1}$ filtered signal vector $\mathbf{x}'_2$ is
\begin{IEEEeqnarray}{rCCCCl}
\mathbf{x}'_2  & = & \mathbf{U}_2^{\text{H}} \mathbf{y}_2 & = & \mathbf{U}_2^{\text{H}} \mathbf{H}_{2,1} \mathbf{V}_1 \mathbf{x}_1 + \mathbf{U}_2^{\text{H}} \mathbf{H}_{2,2} \mathbf{V}_2 \mathbf{x}_2 + \mathbf{U}_2^{\text{H}} \mathbf{n}_2.
\end{IEEEeqnarray}

As the last two column vectors of $\mathbf{H}_{2,1} \mathbf{V}_1$ are the same as the column vectors of $\mathbf{H}_{2,2} \mathbf{V}_2$, we can express filtered signal vector $\mathbf{x}'_2$ as
\begin{equation}
\mathbf{x}'_2 =
  \begin{bmatrix}
x_1^{(1)}\\
x_1^{(2)} + x_2^{(1)}\\
x_1^{(3)} + x_2^{(2)}
  \end{bmatrix} + \mathbf{U}_2^{\text{H}} \mathbf{n}_2.
\end{equation}
Receiver~$2$ then demodulates filtered signal vector $\mathbf{x}'_2$ to network-coded message vector
\begin{equation}
\mathbf{b}'_2 =
  \begin{bmatrix}
b_1^{(1)} \\
b_1^{(2)} \oplus b_2^{(1)}\\
b_1^{(3)} \oplus b_2^{(2)}
  \end{bmatrix}.
\end{equation}
Signal $x_1^{(1)}$ is demodulated by conventional demodulation while the other signals are demodulated by PNC demodulation.

The central processor collects network-coded message vectors $\mathbf{b}'_1$ and $\mathbf{b}'_2$ forwarded from receivers~$1$~and~$2$, respectively, via independent noiseless cooperation links and forms
\begin{equation}
\label{b'_c}
\mathbf{b}'_c =
\begin{bmatrix}
\mathbf{b}'_1 \\
\mathbf{b}'_2
\end{bmatrix}
=
\begin{bmatrix}
b_1^{(1)} \oplus b_2^{(1)}\\
b_1^{(2)} \oplus b_2^{(2)}\\
b_1^{(3)} \\
b_1^{(1)} \\
b_1^{(2)} \oplus b_2^{(1)}\\
b_1^{(3)} \oplus b_2^{(2)}
\end{bmatrix}.
\end{equation}
Obviously, the central processor can recover all original messages of the transmitters by solving any $5$~independent equations with $5$~unknowns in~(\ref{b'_c}). For example, message $b_2^{(1)}$ can be recovered by performing network coding between ${b_1^{(1)} \oplus b_2^{(1)}}$ from $\mathbf{b}'_1$ and $b_1^{(1)}$ from $\mathbf{b}'_2$ such that
\begin{equation}
	(b_1^{(1)} \oplus b_2^{(1)}) \oplus b_1^{(1)} = b_2^{(1)}.
\end{equation} Other original messages of the transmitters can be obtained likewise. As a result, the users can achieve total $d_{\sum} = 5$ DoF over an $N = 3$ symbol extension in time-varying interference channels with limited receiver cooperation by our SNC scheme.

\section{SNC in $K$-User Time-Varying Interference Channels with Limited Receiver Cooperation}
\label{SNC in K-User Time-Varying Interference Channels with Limited Receiver Cooperation}
This section presents the main ideas of this paper. We assume the numbers of the transmitters and the receivers are equal, i.e., ${K = L}$. We show that our SNC scheme can achieve $K$ DoF for the $K$-user time-varying interference channel with limited receiver cooperation (defined in Section~\ref{System Model}), i.e.,
\begin{equation}
\max_{(d_1,d_2,\dots,d_K)} \sum_{k=1}^K d_k = K.
\end{equation}
In other words, full degrees of freedom can be achieved by our SNC scheme.

Here we present our SNC scheme for ${K = 2}$ users. Our scheme for $K \geq 2$ users is provided in Appendix~\ref{SNC scheme for K users}. For ${K=2}$, the system adopts an ${N = n+1}$ symbol extension of the channel where $n \in \mathbb{Z}^{+}$. We show that DoF tuple $(d_1, d_2) = (\frac{n+1}{n+1}, \frac{n}{n+1})$ is achievable.

Our SNC scheme determines the linear precoding matrices of the transmitters so that the receivers obtain independent integer combinations of messages, which are also known as network-coded messages. The central processor then recovers all original messages of the transmitters by solving the linearly independent equations. The details are as follows:

\subsection{Precoding at Transmitters}
First, transmitter~$1$ modulates ${(n+1) \times 1}$ message vector
\begin{equation}
  \mathbf{b}_1 =
  \begin{bmatrix}
b_1^{(1)}&
b_1^{(2)}&
\cdots&
b_1^{(n)}&
b_1^{(n+1)}
  \end{bmatrix}^\text{T}
\end{equation}
to ${(n+1) \times 1}$ signal vector 
\begin{equation}
  \mathbf{x}_1 =
  \begin{bmatrix}
x_1^{(1)}&
x_1^{(2)}&
\cdots&
x_1^{(n)}&
x_1^{(n+1)}
  \end{bmatrix}^\text{T}
\end{equation}
while transmitter~$2$ modulates ${n \times 1}$ message vector
\begin{equation}
  \mathbf{b}_2 =
  \begin{bmatrix}
b_2^{(1)}&
b_2^{(2)}&
\cdots&
b_2^{(n)}
  \end{bmatrix}^\text{T}
\end{equation}
to ${n \times 1}$ signal vector
\begin{equation}
  \mathbf{x}_2 =
  \begin{bmatrix}
x_2^{(1)}&
x_2^{(2)}&
\cdots&
x_2^{(n)}
  \end{bmatrix}^\text{T}.
\end{equation}
Then transmitter~$1$ sends signal vector $\mathbf{x}_{1}$ with ${(n+1) \times (n+1)}$ linear precoding matrix $\mathbf{V}_{1}$ while transmitter~$2$ sends signal vector $\mathbf{x}_{2}$ with ${(n+1) \times n}$ linear precoding matrix $\mathbf{V}_{2}$. 

\subsection{Decoding and Forwarding at Receivers}
The ${(n+1) \times 1}$ received signal vector of receiver ${l \in \{1,2\}}$ is
\begin{equation}
\mathbf{y}_l = \sum_{k=1}^2 \mathbf{H}_{l,k} \mathbf{V}_k \mathbf{x}_k + \mathbf{n}_l,
\end{equation}
where ${(n+1) \times (n+1)}$ channel matrix
\begin{equation}
\mathbf{H}_{l,k} =
  \begin{bmatrix}
{h}_{l,k}^{(1)} & 0 & \dots & 0 \\
0 & {h}_{l,k}^{(2)} & \dots & 0 \\
\vdots & \vdots & \ddots & \vdots\\
0 & 0 & \dots & {h}_{l,k}^{(n+1)}
  \end{bmatrix}.
\end{equation}
$\mathbf{H}_{l,k}$ has full rank ${n+1}$ almost surely because the elements of $\mathbf{H}_{l,k}$ are drawn independently from a continuous distribution.

Our SNC scheme determines the linear precoding matrices of the transmitters so that the receivers obtain independent integer combinations of messages, which are also known as independent network-coded messages. The central processor then recovers all original messages of the transmitters by solving the linear equations. We set up the following signal alignment constraints for the transmitters:
\begin{IEEEeqnarray}{rCl}
\label{1st alignment constraint}
	 \mathbf{H}_{1,2} \mathbf{V}_2 & \prec & \mathbf{H}_{1,1} \mathbf{V}_1, \\
\label{2nd alignment constraint}
	 \mathbf{H}_{2,2} \mathbf{V}_2 & \prec &  \mathbf{H}_{2,1} \mathbf{V}_1
\end{IEEEeqnarray}
where $\mathbf{Q} \prec \mathbf{P}$ denotes that the column vectors of matrix $\mathbf{Q}$ is a subset of those of matrix $\mathbf{P}$ in this paper. In general, for each alignment constraint, the signals are just required to be aligned in the same direction. Taking alignment constraint~(\ref{1st alignment constraint}) as an example, it could be ${\mathbf{H}_{1,2} \mathbf{V}_2  \prec  \alpha \mathbf{H}_{1,1} \mathbf{V}_1}$ where $\alpha$ is a scalar. As we focus on introducing our SNC scheme in this paper, we do not consider the optimization in this aspect.

In order to fulfill alignment constraints~(\ref{1st alignment constraint}) and (\ref{2nd alignment constraint}) and the ideas of SNC, we can set the linear precoding matrices of the transmitters as follows:
\begin{align}
\label{V1}
&\mathbf{V}_1 = 
\begin{bmatrix}
\mathbf{G}_{1,2}^n \mathbf{w} & \mathbf{G}_{1,2}^{n-1} \mathbf{G}_{2,2}\mathbf{w} & \dots & \mathbf{G}_{1,2}\mathbf{G}_{2,2}^{n-1} \mathbf{w} & \mathbf{G}_{2,2}^n\mathbf{w}
\end{bmatrix}, \\
\label{V2}
&\mathbf{V}_2 =
\begin{bmatrix}
\mathbf{G}_{1,2}^{n-1}\mathbf{w} & \mathbf{G}_{1,2}^{n-2}\mathbf{G}_{2,2}\mathbf{w} & \dots & \mathbf{G}_{1,2}\mathbf{G}_{2,2}^{n-2} \mathbf{w} & \mathbf{G}_{2,2}^{n-1}\mathbf{w}
\end{bmatrix}
\end{align}
where $\mathbf{G}_{1,2} = \mathbf{H}_{1,1}^{-1}\mathbf{H}_{1,2}$, $\mathbf{G}_{2,2} = \mathbf{H}_{2,1}^{-1}\mathbf{H}_{2,2}$, and $\mathbf{w}$ is an arbitrary ${(n+1) \times 1}$ column vector. Here $\mathbf{V}_1$ is an ${(n+1) \times (n+1)}$ matrix and $\mathbf{V}_2$ is an ${(n+1) \times n}$ matrix. Without loss of generality, we assume
\begin{equation}
  \mathbf{w} =
  \begin{bmatrix}
1\\
1\\
\vdots\\
1
  \end{bmatrix}.
\end{equation}
As mentioned previously, we focus on introducing our SNC scheme in this paper, therefore we do not consider the optimization of the linear precoding matrices.

\begin{lemma}
\label{V1 and V2}
For ${K=2}$, linear precoding matrix $\mathbf{V}_1$ is a rank ${n+1}$ invertible matrix and linear precoding matrix $\mathbf{V}_2$ has rank~$n$.
\end{lemma}
\begin{IEEEproof}
Let $\text{diag}(\alpha_1, \alpha_2, \dots, \alpha_{n+1}) = \mathbf{G}_{1,2} = \mathbf{H}_{1,1}^{-1}\mathbf{H}_{1,2}$ and $\text{diag}(\beta_1, \beta_2, \dots, \beta_{n+1}) = \mathbf{G}_{2,2} = \mathbf{H}_{2,1}^{-1}\mathbf{H}_{2,2}$. Hence, all $\alpha$ and $\beta$ are independently drawn from a continuous distribution. We express $\mathbf{V}_1$ as
\begin{IEEEeqnarray}{rCl}
\mathbf{V}_1 & = &
\begin{bmatrix}
\alpha_1^n & \alpha_1^{n-1}\beta_1 & \dots  & \beta_1^n \\
\alpha_2^n & \alpha_2^{n-1}\beta_2 & \dots  & \beta_2^n \\
\vdots & \vdots & \ddots & \vdots \\
\alpha_{n+1}^n & \alpha_{n+1}^{n-1}\beta_{n+1} & \dots & \beta_{n+1}^n
\end{bmatrix}.
\end{IEEEeqnarray}
We then left multiply $\mathbf{V}_1$ by ${(n+1) \times (n+1)}$ invertible matrix $\text{diag}(\alpha_1^{-n}, \alpha_2^{-n}, \dots, \alpha_{n+1}^{-n}) = \mathbf{G}_{1,2}^{-n} = (\mathbf{H}_{1,1}^{-1}\mathbf{H}_{1,2})^{-n}$. The result of the multiplication, $\mathbf{G}_{1,2}^{-n}\mathbf{V}_1$, has the same rank as $\mathbf{V}_1$. We express $\mathbf{G}_{1,2}^{-n}\mathbf{V}_1$ as
\begin{equation}
\mathbf{G}_{1,2}^{-n}\mathbf{V}_1  = 
\begin{bmatrix}
1 & \alpha_1^{-1}\beta_1 & (\alpha_1^{-1}\beta_1)^2 & \dots  & (\alpha_1^{-1}\beta_1)^n \\
1 & \alpha_2^{-1}\beta_2 & (\alpha_2^{-1}\beta_2)^2 & \dots & (\alpha_2^{-1}\beta_2)^n \\
\vdots & \vdots & \vdots & \ddots & \vdots \\
1 & \alpha_{n+1}^{-1}\beta_{n+1} & (\alpha_{n+1}^{-1}\beta_{n+1})^2 & \dots & (\alpha_{n+1}^{-1}\beta_{n+1})^n
\end{bmatrix},
\end{equation}
which is a Vandermonde matrix. Hence, the determinant of $\mathbf{G}_{1,2}^{-n}\mathbf{V}_1$ is
\begin{equation}
\text{det}(\mathbf{G}_{1,2}^{-n}\mathbf{V}_1) = \prod_{1 \leq i < j \leq n+1} (\alpha_j^{-1}\beta_j - \alpha_i^{-1}\beta_i).
\end{equation}
As all $\alpha$ and $\beta$ are independently drawn from a continuous distribution, $\text{det}(\mathbf{G}_{1,2}^{-n}\mathbf{V}_1)$ is non-zero almost surely. Therefore, $\mathbf{G}_{1,2}^{-n}\mathbf{V}_1$ and $\mathbf{V}_1$ have rank~${n+1}$.

The matrix formed by the first $n$~rows of $\mathbf{V}_2$ has a similar structure to $\mathbf{V}_1$, therefore $\mathbf{V}_2$ can be proved to have rank~$n$ in a similar way. The proof is omitted to avoid repetition. Moreover, an alternative proof of Lemma~\ref{V1 and V2} is presented in Appendix~\ref{Alternative proof}. This alternative proof is useful later in this paper.
\end{IEEEproof}

Now we show that the signals from the transmitters are aligned at receiver~$1$. The multiplications of the channel matrices and the linear precoding matrices at receiver~$1$ are
\begin{align}
\label{H11V1}
&\mathbf{H}_{1,1} \mathbf{V}_1 \notag \\ = \ &[\mathbf{H}_{1,2}\mathbf{G}_{1,2}^{n-1}\mathbf{w}, \mathbf{H}_{1,2} \mathbf{G}_{1,2}^{n-2}\mathbf{G}_{2,2} \mathbf{w}, \dots, \mathbf{H}_{1,2} \mathbf{G}_{2,2}^{n-1} \mathbf{w}, \mathbf{H}_{1,1} \mathbf{G}_{2,2}^n \mathbf{w}], \\
\label{H12V2}
&\mathbf{H}_{1,2} \mathbf{V}_2 \notag \\ = \ &[\mathbf{H}_{1,2}\mathbf{G}_{1,2}^{n-1}\mathbf{w}, \mathbf{H}_{1,2}\mathbf{G}_{1,2}^{n-2}\mathbf{G}_{2,2}\mathbf{w}, \dots, \mathbf{H}_{1,2}\mathbf{G}_{2,2}^{n-1}\mathbf{w}],
\end{align}
where $\mathbf{H}_{1,1}\mathbf{V}_1$ is an ${(n+1) \times (n+1)}$ matrix and $\mathbf{H}_{1,2} \mathbf{V}_2$ is an ${(n+1) \times n}$ matrix. The column vectors in (\ref{H11V1}) and (\ref{H12V2}) are separated by commas due to space limitation. The first $n$~column vectors of $\mathbf{H}_{1,1} \mathbf{V}_1$ are the same as the column vectors of $\mathbf{H}_{1,2} \mathbf{V}_2$, therefore signal alignment constraint~(\ref{1st alignment constraint}) is satisfied.

As both $\mathbf{H}_{1,1}$ and $\mathbf{V}_1$ have full rank, receiver~$1$ applies ${(n+1) \times (n+1)}$ linear filtering matrix $\mathbf{U}_1^{\text{H}} = (\mathbf{H}_{1,1} \mathbf{V}_1)^{-1}$ to $(n+1) \times 1$ received signal vector $\mathbf{y}_1$. The ${(n+1) \times 1}$ filtered signal vector $\mathbf{x}'_1$ is
\begin{IEEEeqnarray}{rCl}
\mathbf{x}'_1  & = & \mathbf{U}_1^{\text{H}} \mathbf{y}_1 \nonumber \\ & = & \sum_{k=1}^2 \mathbf{U}_1^{\text{H}} \mathbf{H}_{1,k} \mathbf{V}_k \mathbf{x}_k + \mathbf{U}_1^{\text{H}} \mathbf{n}_1.
\end{IEEEeqnarray}
As the first $n$~column vectors of $\mathbf{H}_{1,1} \mathbf{V}_1$ are the same as the column vectors of $\mathbf{H}_{1,2} \mathbf{V}_2$, we can express filtered signal vector $\mathbf{x}'_1$ as
\begin{IEEEeqnarray}{rCl}
\label{x'_1}
\mathbf{x}'_1 & = &
	\begin{bmatrix}
		\mathbf{U}_1 \mathbf{H}_{1,1} \mathbf{V}_1 & \mathbf{U}_1 \mathbf{H}_{1,2} \mathbf{V}_2
	\end{bmatrix}
    \begin{bmatrix}
    	\mathbf{x}_1 \\ \mathbf{x}_2
  	\end{bmatrix} + \mathbf{U}_1^{\text{H}} \mathbf{n}_1 \nonumber \\ & = &
  \begin{bmatrix}
x_1^{(1)} + x_2^{(1)}\\
x_1^{(2)} + x_2^{(2)}\\
\vdots\\
x_1^{(n)} + x_2^{(n)}\\
x_1^{(n+1)}
  \end{bmatrix} + \mathbf{U}_1^{\text{H}} \mathbf{n}_1.
\end{IEEEeqnarray} Here ${(n+1) \times (2n+1)}$ effective channel  matrix
$\begin{bmatrix}
	\mathbf{U}_1 \mathbf{H}_{1,1} \mathbf{V}_1 & \mathbf{U}_1 \mathbf{H}_{1,2} \mathbf{V}_2
\end{bmatrix}$
is a binary matrix and it affects the alignment of the signals at receiver~$1$. It is worth noting that there are no non-integer parts in the effective channel matrix. Receiver~$1$ then demodulates filtered signal vector $\mathbf{x}'_1$ to ${(n+1) \times 1}$ network-coded message vector $\mathbf{b}'_1$ over $\text{GF}(q)$ where
\begin{IEEEeqnarray}{rCl}
\label{b'_1}
\mathbf{b}'_1 & = &
	\begin{bmatrix}
		\mathbf{U}_1 \mathbf{H}_{1,1} \mathbf{V}_1 & \mathbf{U}_1 \mathbf{H}_{1,2} \mathbf{V}_2
	\end{bmatrix}
    \begin{bmatrix}
    	\mathbf{b}_1 \\ \mathbf{b}_2
  	\end{bmatrix} \nonumber \\ & = &
  \begin{bmatrix}
b_1^{(1)} \oplus b_2^{(1)}\\
b_1^{(2)} \oplus b_2^{(2)}\\
\vdots\\
b_1^{(n)} \oplus b_2^{(n)}\\
b_1^{(n+1)}
  \end{bmatrix}.
\end{IEEEeqnarray}
Signal $x_1^{(n+1)}$ is demodulated by conventional demodulation while the other signals are demodulated by PNC demodulation.

The messages decoded at receiver~$2$ can be understood likewise. The multiplications of the channel matrices and the linear precoding matrices at receiver~$2$ are
\begin{align}
\label{H21V1}
&\mathbf{H}_{2,1} \mathbf{V}_1 \notag \\ = \ &[\mathbf{H}_{2,1} \mathbf{G}_{1,2}^n \mathbf{w}, \mathbf{H}_{2,2} \mathbf{G}_{1,2}^{n-1} \mathbf{w}, \dots, \mathbf{H}_{2,2} \mathbf{G}_{1,2} \mathbf{G}_{2,2}^{n-2} \mathbf{w}, \mathbf{H}_{2,2} \mathbf{G}_{2,2}^{n-1} \mathbf{w}], \\
\label{H22V2}
&\mathbf{H}_{2,2} \mathbf{V}_2 \notag \\ = \ &[\mathbf{H}_{2,2} \mathbf{G}_{1,2}^{n-1} \mathbf{w}, \dots, \mathbf{H}_{2,2} \mathbf{G}_{1,2} \mathbf{G}_{2,2}^{n-2}\mathbf{w}, \mathbf{H}_{2,2} \mathbf{G}_{2,2}^{n-1} \mathbf{w}],
\end{align}
where $\mathbf{H}_{2,1} \mathbf{V}_1$ is an ${(n+1) \times (n+1)}$ matrix and $\mathbf{H}_{2,2} \mathbf{V}_2$ is an ${(n+1) \times n}$ matrix. The column vectors in (\ref{H21V1}) and (\ref{H22V2}) are separated by commas. The last $n$~column vectors of $\mathbf{H}_{2,1} \mathbf{V}_1$ are the same as the column vectors of $\mathbf{H}_{2,2} \mathbf{V}_2$, therefore signal alignment constraint~(\ref{2nd alignment constraint}) is satisfied.

Receiver~$2$ applies ${(n+1) \times (n+1)}$ linear filtering matrix $\mathbf{U}_2^{\text{H}} = (\mathbf{H}_{2,1} \mathbf{V}_1)^{-1}$ to ${(n+1) \times 1}$ received signal vector $\mathbf{y}_2$. The ${(n+1) \times 1}$ filtered signal vector $\mathbf{x}'_2$ is
\begin{IEEEeqnarray}{rCl}
\mathbf{x}'_2  & = & \mathbf{U}_2^{\text{H}} \mathbf{y}_2 \nonumber \\ & = & \sum_{k=1}^2 \mathbf{U}_2^{\text{H}} \mathbf{H}_{2,k} \mathbf{V}_k \mathbf{x}_k + \mathbf{U}_2^{\text{H}} \mathbf{n}_2.
\end{IEEEeqnarray}
As the last $n$~column vectors of $\mathbf{H}_{2,1} \mathbf{V}_1$ are the same as the column vectors of $\mathbf{H}_{2,2} \mathbf{V}_2$, we can express ${(n+1) \times 1}$ filtered signal vector $\mathbf{x}'_2$ as
\begin{IEEEeqnarray}{rCl}
\label{x'_2}
\mathbf{x}'_2 & = &
	\begin{bmatrix}
		\mathbf{U}_2 \mathbf{H}_{2,1} \mathbf{V}_1 & \mathbf{U}_2 \mathbf{H}_{2,2} \mathbf{V}_2
	\end{bmatrix}
    \begin{bmatrix}
    	\mathbf{x}_1 \\ \mathbf{x}_2
  	\end{bmatrix} + \mathbf{U}_2^{\text{H}} \mathbf{n}_2 \nonumber \\ & = &
  \begin{bmatrix}
x_1^{(1)}\\
x_1^{(2)} + x_2^{(1)}\\
\vdots\\
x_1^{(n)} + x_2^{(n-1)}\\
x_1^{(n+1)} + x_2^{(n)}
  \end{bmatrix} + \mathbf{U}_2^{\text{H}} \mathbf{n}_2.
\end{IEEEeqnarray} Here ${(n+1) \times (2n+1)}$ effective channel matrix
$\begin{bmatrix}
	\mathbf{U}_2 \mathbf{H}_{2,1} \mathbf{V}_1 & \mathbf{U}_2 \mathbf{H}_{2,2} \mathbf{V}_2
\end{bmatrix}$
, which is a binary matrix, affects the alignment of the signals at receiver~$2$. Filtered signal vector $\mathbf{x}'_2$ is then demodulated to ${(n+1) \times 1}$ network-coded message vector $\mathbf{b}'_2$ over $\text{GF}(q)$ where
\begin{IEEEeqnarray}{rCl}
\label{b'_2}
\mathbf{b}'_2 & = &
	\begin{bmatrix}
		\mathbf{U}_2 \mathbf{H}_{2,1} \mathbf{V}_1 & \mathbf{U}_2 \mathbf{H}_{2,2} \mathbf{V}_2
	\end{bmatrix}
    \begin{bmatrix}
    	\mathbf{b}_1 \\ \mathbf{b}_2
  	\end{bmatrix} \nonumber \\ & = &
  \begin{bmatrix}
b_1^{(1)} \\
b_1^{(2)} \oplus b_2^{(1)}\\
\vdots\\
b_1^{(n)} \oplus b_2^{(n-1)}\\
b_1^{(n+1)} \oplus b_2^{(n)}
  \end{bmatrix}.
\end{IEEEeqnarray}
Signal $x_1^{(1)}$ is demodulated by conventional demodulation while the other signals are demodulated by PNC demodulation.

\subsection{Decoding at Central Processor}
We consider the wired cooperation links only support forwarding the digital decoded packets, rather than analog signal samples, from the receivers to the central processor. The central processor collects network-coded message vectors $\mathbf{b}'_1$ and $\mathbf{b}'_2$ forwarded from receivers~$1$~and~$2$, respectively, via independent noiseless cooperation links. The central processor forms ${(2n+2) \times 1}$ network-coded message vector $\mathbf{b}'_c$ over $\text{GF}(q)$ where
\begin{IEEEeqnarray}{rCl}
\mathbf{b}'_c & = &
\begin{bmatrix}
\mathbf{b}'_1 \\ \mathbf{b}'_2
\end{bmatrix} \nonumber \\
& = &
\begin{bmatrix}
		\mathbf{U}_1 \mathbf{H}_{1,1} \mathbf{V}_1 & \mathbf{U}_1 \mathbf{H}_{1,2} \mathbf{V}_2 \\
				\mathbf{U}_2 \mathbf{H}_{2,1} \mathbf{V}_1 & \mathbf{U}_2 \mathbf{H}_{2,2} \mathbf{V}_2
\end{bmatrix}
\begin{bmatrix}
\mathbf{b}_1 \\ \mathbf{b}_2
\end{bmatrix}.
\end{IEEEeqnarray}
Let ${(2n+2) \times (2n+1)}$ binary matrix
\begin{equation}
\mathbf{F} = 
\begin{bmatrix}
		\mathbf{U}_1 \mathbf{H}_{1,1} \mathbf{V}_1 & \mathbf{U}_1 \mathbf{H}_{1,2} \mathbf{V}_2 \\
				\mathbf{U}_2 \mathbf{H}_{2,1} \mathbf{V}_1 & \mathbf{U}_2 \mathbf{H}_{2,2} \mathbf{V}_2
\end{bmatrix}
\end{equation}
over $\text{GF}(q)$.

\begin{lemma}
\label{F has rank 2n+1} 
For ${K = 2}$, matrix $\mathbf{F}$ has full rank~${2n+1}$ for some finite field sizes~$q$.
\end{lemma}
\begin{IEEEproof}
We first assume $\mathbf{F}$ is in the field of complex numbers. We can decompose $\mathbf{F}$ as
\begin{equation}
\mathbf{F} = 
\begin{bmatrix}
\mathbf{U}_1 & 0 \\
0 & \mathbf{U}_2
\end{bmatrix}
\begin{bmatrix}
		 \mathbf{H}_{1,1} & \mathbf{H}_{1,2}  \\
				 \mathbf{H}_{2,1}  &  \mathbf{H}_{2,2} 
\end{bmatrix}
\begin{bmatrix}
\mathbf{V}_1 & 0 \\
0 & \mathbf{V}_2
\end{bmatrix}.
\end{equation}
As we have mentioned before, ${(n+1) \times (n+1)}$ matrices $\mathbf{U}_1$, $\mathbf{U}_2$, and $\mathbf{V}_1$ have rank~$n+1$ while ${(n+1) \times n}$ matrix $\mathbf{V}_2$ has rank~$n$. Therefore, ${(2n+2) \times (2n+2)}$ matrix
$\begin{bmatrix}
\mathbf{U}_1 & 0 \\
0 & \mathbf{U}_2
\end{bmatrix}$
has rank~${2n+2}$ and $(2n+2) \times (2n+1)$ matrix
$\begin{bmatrix}
\mathbf{V}_1 & 0 \\
0 & \mathbf{V}_2
\end{bmatrix}$
has rank~${2n+1}$. Moreover, as all channel coefficients are independently drawn from a continuous distribution,  ${(2n+2) \times (2n+2)}$ matrix
$\begin{bmatrix}
		 \mathbf{H}_{1,1} & \mathbf{H}_{1,2}  \\
				 \mathbf{H}_{2,1}  &  \mathbf{H}_{2,2} 
\end{bmatrix}$ has rank~$2n+2$ almost surely. As matrices
$\begin{bmatrix}
\mathbf{U}_1 & 0 \\
0 & \mathbf{U}_2
\end{bmatrix}$
and
$\begin{bmatrix}
\mathbf{H}_{1,1} & \mathbf{H}_{1,2}  \\
\mathbf{H}_{2,1}  &  \mathbf{H}_{2,2} 
\end{bmatrix}$
are invertible, matrix $\mathbf{F}$ should has the same rank as matrix 
$\begin{bmatrix}
\mathbf{V}_1 & 0 \\
0 & \mathbf{V}_2
\end{bmatrix}$. Hence, binary matrix $\mathbf{F}$ has rank~$2n+1$ in the field of complex numbers. 

We pick ${2n+1}$~independent rows of binary matrix $\mathbf{F}$ to form a ${(2n+1) \times (2n+1)}$ invertible binary matrix $\mathbf{F}'$, where $\text{det}(\mathbf{F}') \neq 0$. As there must exist some~$q$ where $\text{det}(\mathbf{F}') \ \text{mod} \ q \neq 0$, binary matrix $\mathbf{F}$ has rank~$2n+1$ over $\text{GF}(q)$ for some finite field sizes~$q$.
\end{IEEEproof}

In other words, there are ${2n+1}$ independent equations (integer combinations of original messages) with ${2n+1}$ unknowns (messages) at the central processor. The central processor can recover all original messages of the transmitters by solving the equations. Hence, the system achieves total ${2n+1}$ DoF over an ${N = n+1}$ symbol extension for any positive integer~$n$ by our SNC scheme.

The general SNC scheme for $K \geq 2$ users is presented in Appendix~\ref{SNC scheme for K users}. Our SNC scheme for general MIMO cases with arbitrary numbers of transmitters, receivers, and antennas of each node can be found in Appendix~\ref{Degrees of Freedom for MIMO General Cases}. Notice that the achievement of full DoF in our SNC scheme is not limited by the number of antennas of each node.

\subsection{End-to-End Sum-Rate}
In the SNC scheme, the signals from each transmitter are decoded by more than one receivers. Therefore, the achievable rate of sending signal $x_{k}^{(n)}$ from transmitter $k$ is
\begin{IEEEeqnarray}{rCl}
R_{k}^{(n)}  = \min_{l \in \{1,2,\dots,K)} {R}_{l,k}^{(n)},
\end{IEEEeqnarray} where ${R}_{l,k}^{(n)}$ is the achievable rate of sending signal $x_{k}^{(n)}$ from transmitter $k$ to receiver $l$ defined in Section~\ref{System Model}. Finally, the end-to-end sum-rate in an $N$ symbol extension is
\begin{equation}
	R_{\text{sum}} = \sum_{k=1}^{K} \sum_{n=1}^{N} \min(R_{k}^{(n)},\log_2(1+\bar{\rho})).
\end{equation}
As defined in Section~\ref{System Model}, $\log_2(1+\bar{\rho})$ is the restricted capacity of each cooperation link and $\bar{\rho}$ is the average SNR of the links in the interference channel.

\section{Simulation Results}
\label{Simulation Results}
\begin{figure}[!t]
\centering
   \includegraphics[width=3.5in]{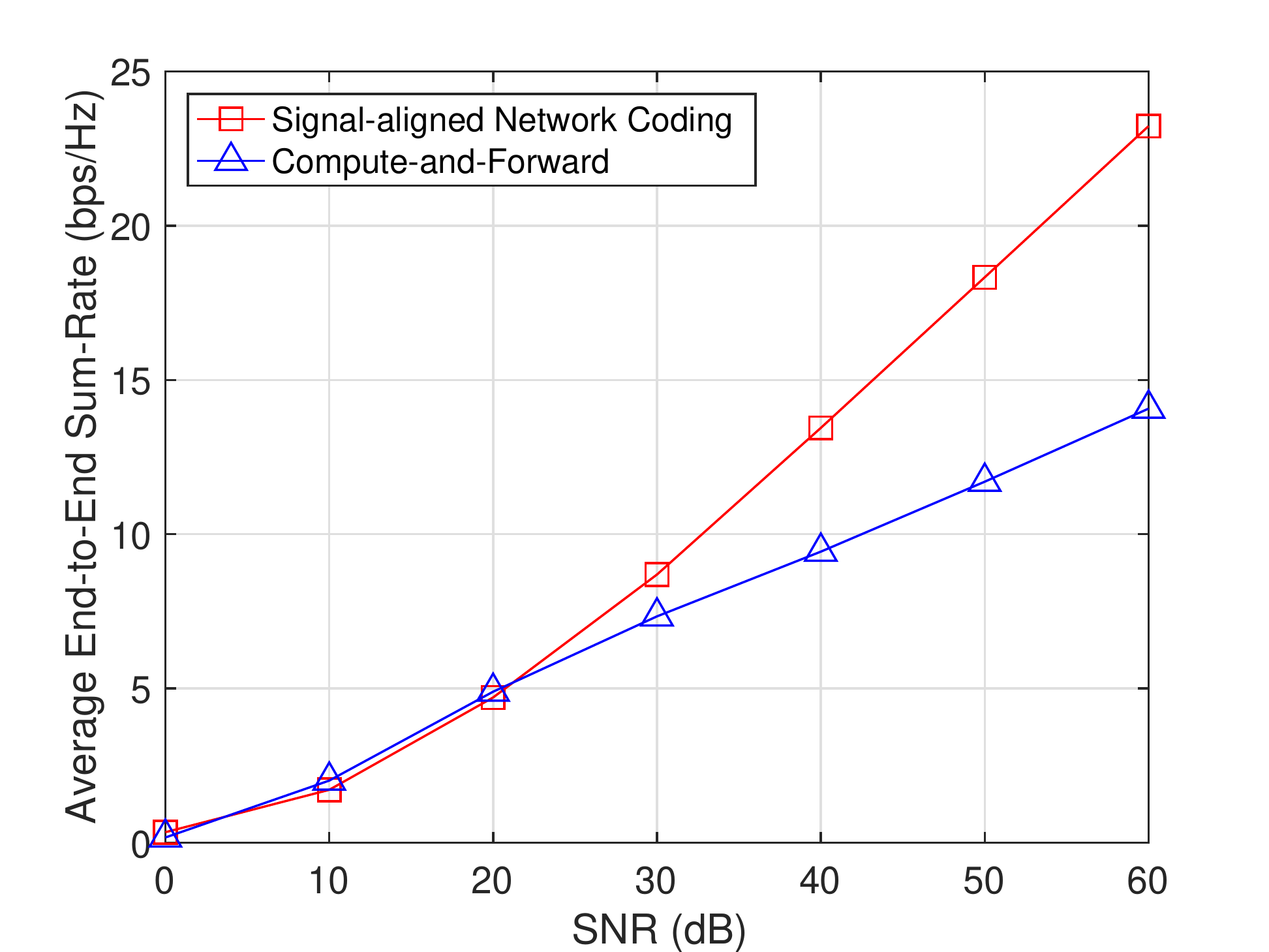}
  \caption{Average end-to-end sum-rate achieved in two-user time-varying interference channels with limited receiver cooperation.}
  \label{two-user sum-rate}
\end{figure}
\begin{figure}[!t]
\centering
   \includegraphics[width=3.5in]{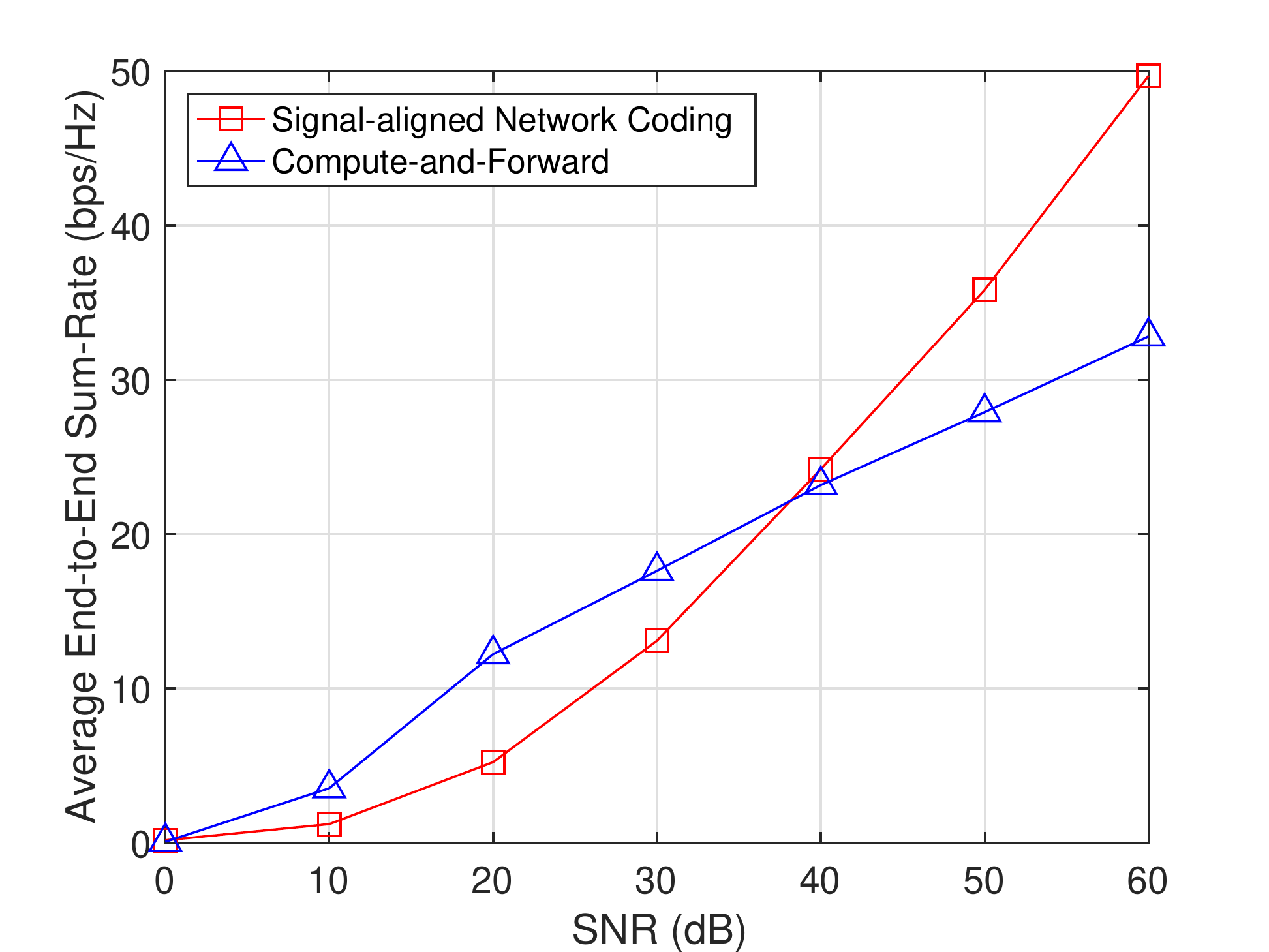}
  \caption{Average end-to-end sum-rate achieved in three-user time-varying interference channels with limited receiver cooperation.}
  \label{three-user sum-rate}
\end{figure}
We present the simulation results to evaluate the end-to-end sum-rate performance of the SNC scheme in time-varying interference channels with limited receiver cooperation. The performance of our SNC scheme is compared with that of the compute-and-forward scheme. As the selection of the optimal integer-valued equation coefficient vectors in the compute-and-forward scheme turns out to be a shortest vector problem (SVP) which is NP-hard, we adopt the quantization algorithm proposed in~\cite{A quadratic programming relaxation approach to compute-and-forward network coding design} for finding a suboptimal equation coefficient vector. This quantization algorithm considers the real-value channel model, and hence we simulate the performances of the schemes with real-value channel coefficients in this section.

We assume the transmit powers for transmitting signals in all nodes are the same and the noise variances at each node are the same. We consider that all systems adopt the binary phase shift keying (BPSK) modulation. Simulation results are illustrated with respect to the ratio of transmit power for each signal to noise variance at each node (i.e., SNR value) from 0 to 60 dB. The end-to-end sum-rate are computed using 1000 random channel realizations. The obtained results for the two-user and the three-user cases with one antenna per node are shown in {\figurename~\ref{two-user sum-rate} and \figurename~\ref{three-user sum-rate}} respectively.

Considering the finite SNR performance, {\figurename~\ref{two-user sum-rate}--\ref{three-user sum-rate}} show that our SNC scheme outperforms the compute-and-forward scheme in the two-user and the three-user cases, especially in the high SNR regime. The sum-rate of the compute-and-forward scheme is deteriorated by the non-integer penalty and the rank deficiency described in Section~\ref{Introduction}. The performance improvement of our SNC scheme mainly comes from efficient utilization of the signal subspaces for conveying independent integer combinations of messages to the central processor.

\section{Conclusion}
\label{Conclusion}
In this paper, we propose a SNC scheme in $K$-user time-varying MIMO interference channels with limited receiver cooperation. This channel model widely characterizes the scenarios in C-RAN, distributed MIMO, WLAN, etc. Our SNC scheme utilizes the precoding matrices of the transmitters to make different integer combinations of the transmitted messages, which are linearly independent, at each receiver. We prove that our scheme is able to approach arbitrarily close to the upper DoF bound. The DoF performance improvement of our scheme mainly comes from efficient utilization of the signal subspaces for conveying linearly independent equations of messages. This improvement is not limited by the number of antennas of each node. In terms of end-to-end sum-rate, simulation results show that our SNC scheme achieves superior performance compared to the compute-and-forward scheme in the two-user and the three-user cases with BPSK modulation scheme. Finally, sum-rate maximization by optimizing the precoding matrices in our scheme remains an open problem. This could be considered as a future work.


%

\appendices
\section{General SNC scheme for $K \geq 2$ users}
\label{SNC scheme for K users}
We denote ${N = \binom{n+K(K-1)-1}{n}}$ and ${N' = \binom{n+K(K-1)-2}{n-1}}$ where $n \in \mathbb{Z}^{+}$. The system adopts an $N$ symbol extension for ${K \geq 2}$. We use transmitter~${k' \in \{2,3,\dots,K\}}$ to denote each transmitter except transmitter~$1$. We also use receiver~${l' \in \{2,3,\dots,K\}}$ to denote each receiver except receiver~$1$. We show that degrees of freedom tuple $(d_1, d_2, \dots, d_K)$ is achievable where
\begin{IEEEeqnarray}{rCl}
d_1 & = & \frac{N}{N} = 1, \\
d_{k'} & = & \frac{N'}{N} = \frac{n}{n+K(K-1)-1}, \forall k' \in \{2, 3, \dots, K\}.
\end{IEEEeqnarray}
In other words, 
\begin{equation}
\label{achievable degrees of freedom tuple for K users}
\sup_n \frac{N+(K-1)N'}{N} = \sup_n \frac{Kn+K(K-1)-1}{n+K(K-1)-1} = K.\
\end{equation}

First of all, transmitter~$1$ modulates ${N \times 1}$ message vector $\mathbf{b}_1$ to ${N \times 1}$ signal vector $\mathbf{x}_1$ while transmitter~$k'$ modulates ${N' \times 1}$ message vector $\mathbf{b}_{k'}$ to ${N' \times 1}$ signal vector $\mathbf{x}_{k'}$. Transmitter~${k \in \{1,2,\dots,K\}}$ sends signal vector $\mathbf{x}_{k}$ with linear precoding matrix $\mathbf{V}_{k}$. The ${N \times 1}$ received signal vector at receiver~${l \in \{1,2,\dots,K\}}$ is
\begin{equation}
\mathbf{y}_l = \sum_{k=1}^K \mathbf{H}_{l,k} \mathbf{V}_k \mathbf{x}_k + \mathbf{n}_l,
\end{equation}
where ${N \times N}$ channel matrix
\begin{equation}
\mathbf{H}_{l,k} =
  \begin{bmatrix}
{h}_{l,k}^{(1)} & 0 & \dots & 0 \\
0 & {h}_{l,k}^{(2)} & \dots & 0 \\
\vdots & \vdots & \ddots & \vdots\\
0 & 0 & \dots & {h}_{l,k}^{(N)}
  \end{bmatrix}.
\end{equation}

We set up the following signal alignment constraints for the transmitters which are similar to the ${K = 2}$ case:
\begin{IEEEeqnarray}{rCCl}
\label{alignment constraint for K user}
	  \mathbf{H}_{l,k'} \mathbf{V}_{k'} & \prec & \mathbf{H}_{l,1} \mathbf{V}_1, & \forall l \in \{1, 2, \dots, K\}, \nonumber \\ & & & \forall k' \in \{2, 3, \dots, K\},
\end{IEEEeqnarray}
where $\mathbf{Q} \prec \mathbf{P}$ denotes the column vectors of matrix $\mathbf{Q}$ is a subset of those of matrix $\mathbf{P}$. In order to fulfill alignment constraints~(\ref{alignment constraint for K user}) and the ideas of SNC, we can set up the linear precoding matrices of the transmitters according to the following way. The sets of column vectors of $\mathbf{V}_1$ and $\mathbf{V}_{k'}$ are equal to the sets $V_1$ and $V_{k'}$ respectively. Here
\begin{align}
V_1 = &\left\{ \left(\prod_{\substack{i \in \{1,2,\dots,K\},\\ j \in \{2,3,\dots,K\}}} \mathbf{G}_{i,j}^{n_{i,j}} \right) \mathbf{w}: \sum_{i,j} n_{i,j} = n, n_{i,j} \in \mathbb{Z}^{+} \right\}
, \\ 
V_{k'} = &\left\{ \left(\prod_{\substack{i \in \{1,2,\dots,K\},\\ j \in \{2,3,\dots,K\}}} \mathbf{G}_{i,j}^{n_{i,j}} \right) \mathbf{w}: \sum_{i,j} n_{i,j} = n-1, n_{i,j} \in \mathbb{Z}^{+} \right\}
, \nonumber \\
& \ \forall k' \in \{2,3,\dots,K\}
\end{align}
where $\mathbf{G}_{i,j} = \mathbf{H}_{i,1}^{-1}\mathbf{H}_{i,j}$ and $\mathbf{w}$ is an arbitrary ${N \times 1}$ column vector. Without loss of generality, we assume
\begin{equation}
  \mathbf{w} =
  \begin{bmatrix}
1&
1&
\cdots&
1
  \end{bmatrix}^{\text{T}}.
\end{equation}
Here $\mathbf{V}_1$ is an ${N \times N}$ matrix and $\mathbf{V}_{k'}$ are ${N \times N'}$ matrices. The signals from the transmitters are aligned at the receivers. We take receiver~$1$ as an example. The column vectors of matrix $\mathbf{H}_{1,k'} \mathbf{V}_{k'}$ is a subset of those of matrix $\mathbf{H}_{1,1} \mathbf{V}_{1}$. Hence, the signals from all of the transmitters are aligned at receiver~$1$. Likewise, considering any column vector in $\mathbf{H}_{l,k'} \mathbf{V}_{k'}$, it can be found in $\mathbf{H}_{l,1} \mathbf{V}_{1}$. 

Linear precoding matrices $\mathbf{V}_1$ and $\mathbf{V}_{k'}$ have ranks~$N$~and~$N'$ respectively. An iterative argument which is similar to that in Appendix~\ref{Alternative proof} for the $K = 2$ case can be used. Here we just briefly describe the main ideas of the proofs to avoid repetition. We can proof ${\text{det}(\mathbf{V}_1) \neq 0}$ by contradiction. We expand the determinant of $\mathbf{V}_1$ along the first row and then keep deducing the determinant of the submatrix formed by eliminating the first row and column is required to be zero repeatedly. Finally, it remains an entry of which value is drawn from a continuous distribution, and hence ${\text{det}(\mathbf{V}_1) \neq 0}$. In other words, $\mathbf{V}_1$ has rank~$N$. As the matrix formed by the first $N'$~rows of $\mathbf{V}_{k'}$ has a similar structure to $\mathbf{V}_1$, $\mathbf{V}_{k'}$ can be proved to have rank~$N'$ likewise.

Matrix $\mathbf{H}_{l,1} \mathbf{V}_1$ is invertible almost certainly as the elements of channel matrices $\mathbf{H}_{l,1}$ are drawn independently from a continuous distribution and $\mathbf{V}_1$ is invertible. Receiver~$l$ applies ${N \times N}$ linear filtering matrix $\mathbf{U}_l^{\text{H}} = (\mathbf{H}_{l,1} \mathbf{V}_1)^{-1}$ to ${N \times 1}$ received signal vector $\mathbf{y}_l$ such that
\begin{IEEEeqnarray}{rCl}
\mathbf{x}'_l  & = & \mathbf{U}_l^{\text{H}} \mathbf{y}_l \nonumber \\ & = & \sum_{k=1}^K \mathbf{U}_l \mathbf{H}_{l,k} \mathbf{V}_k \mathbf{x}_k + \mathbf{U}_l \mathbf{n}_l.
\end{IEEEeqnarray}
Filtered signal vector $\mathbf{x}'_l$ is then demodulated to network-coded message $\mathbf{b}'_l$. The aligned signals are demodulated by PNC demodulation while the other signals are demodulated by conventional demodulation. 

The central processor pools network-coded message vectors $\mathbf{b}'_l$ forwarded from receiver~$l$ via independent noiseless cooperation links. There exists ${N+(K-1)N'}$ independent integer combinations of original messages with ${N+(K-1)N'}$ unknown messages at the central processor. The proof is similar to that in Lemma~\ref{F has rank 2n+1}. The central processor forms ${KN \times 1}$ network-coded message vector $\mathbf{b}'_c$ over $\text{GF}(q)$ where
\begin{equation}
\mathbf{b}'_c =
\begin{bmatrix}
\mathbf{U}_1 \mathbf{H}_{1,1} \mathbf{V}_1 & \mathbf{U}_1 \mathbf{H}_{1,2} \mathbf{V}_2 & \dots & \mathbf{U}_1 \mathbf{H}_{1,K} \mathbf{V}_K \\
\mathbf{U}_2 \mathbf{H}_{2,1} \mathbf{V}_1 & \mathbf{U}_2 \mathbf{H}_{2,2} \mathbf{V}_2 & \dots & \mathbf{U}_2 \mathbf{H}_{2,K} \mathbf{V}_K \\
\vdots & \vdots & \ddots & \vdots \\
\mathbf{U}_K \mathbf{H}_{K,1} \mathbf{V}_1 & \mathbf{U}_K \mathbf{H}_{K,2} \mathbf{V}_2 & \dots & \mathbf{U}_K \mathbf{H}_{K,K} \mathbf{V}_K
\end{bmatrix}
\begin{bmatrix}
\mathbf{b}_1 \\
\mathbf{b}_2 \\
\vdots \\
\mathbf{b}_K
\end{bmatrix}.
\end{equation}
Let ${KN \times [N+(K-1)N']}$ binary matrix
\begin{equation}
\mathbf{F} = 
\begin{bmatrix}
\mathbf{U}_1 \mathbf{H}_{1,1} \mathbf{V}_1 & \mathbf{U}_1 \mathbf{H}_{1,2} \mathbf{V}_2 & \dots & \mathbf{U}_1 \mathbf{H}_{1,K} \mathbf{V}_K \\
\mathbf{U}_2 \mathbf{H}_{2,1} \mathbf{V}_1 & \mathbf{U}_2 \mathbf{H}_{2,2} \mathbf{V}_2 & \dots & \mathbf{U}_2 \mathbf{H}_{2,K} \mathbf{V}_K \\
\vdots & \vdots & \ddots & \vdots \\
\mathbf{U}_K \mathbf{H}_{K,1} \mathbf{V}_1 & \mathbf{U}_K \mathbf{H}_{K,2} \mathbf{V}_2 & \dots & \mathbf{U}_K \mathbf{H}_{K,K} \mathbf{V}_K
\end{bmatrix}
\end{equation}
over $\text{GF}(q)$. We first assume $\mathbf{F}$ is in the field of complex numbers and decompose it into matrix multiplications
$\begin{bmatrix}
\mathbf{U}
\end{bmatrix}
\begin{bmatrix}
\mathbf{H}
\end{bmatrix}
\begin{bmatrix}
\mathbf{V}
\end{bmatrix}$ where
\begin{align}
\begin{bmatrix}
\mathbf{U}
\end{bmatrix} &=
\begin{bmatrix}
\mathbf{U}_1 & 0 & \dots & 0 \\
0 & \mathbf{U}_2 & \dots & 0 \\
\vdots & \vdots & \ddots & \vdots \\
0 & 0 & \dots & \mathbf{U}_K
\end{bmatrix}, \\
\begin{bmatrix}
\mathbf{H}
\end{bmatrix} &=
\begin{bmatrix}
\mathbf{H}_{1,1} & \mathbf{H}_{1,2} & \dots & \mathbf{H}_{1,K} \\
\mathbf{H}_{2,1}  &  \mathbf{H}_{2,2} & \dots & \mathbf{H}_{2,K} \\
\vdots & \vdots & \ddots & \vdots \\
\mathbf{H}_{K,1}  &  \mathbf{H}_{K,2} & \dots & \mathbf{H}_{K,K}
\end{bmatrix}, \\
\begin{bmatrix}
\mathbf{V}
\end{bmatrix} &=
\begin{bmatrix}
\mathbf{V}_1 & 0 & \dots & 0 \\
0 & \mathbf{V}_2 & \dots & 0 \\
\vdots & \vdots & \ddots & \vdots \\
0 & 0 & \dots & \mathbf{V}_K
\end{bmatrix}. 
\end{align}
As mentioned before, ${N \times N}$ matrices $\mathbf{U}_l$ and $\mathbf{V}_1$ have rank $N$. Moreover, ${N \times N'}$ matrices $\mathbf{V}_k'$ have rank $N'$. Recall that all channel coefficients are drawn from a continuous distribution. Therefore, $\mathbf{F}$ has rank ${N+(K-1)N'}$. We pick ${N+(K-1)N'}$ independent rows of $\mathbf{F}$ to form an ${(N+(K-1)N') \times (N+(K-1)N')}$ invertible binary matrix $\mathbf{F'}$, which means $\text{det}(\mathbf{F'}) \neq 0$. As there must exist some $q$ such that $\text{det}(\mathbf{F'}) \ \text{mod} \ q \neq 0$, binary matrix $\mathbf{F}$ has rank ${N+(K-1)N'}$ over $\text{GF}(q)$ with some finite field sizes~$q$. 

In other words, there are ${N+(K-1)N'}$ independent linear equations with ${N+(K-1)N'}$ unknowns collected by the central processor. The central processor can recover all original messages of the transmitters by solving the equations. Therefore, the system achieves total ${N+(K-1)N'}$ DoF over an $N$ symbol extension by our SNC scheme. Considering $n$ tends to infinity, $K$~DoF can be achieved by our SNC scheme, equivalent to unlimited cooperation at the receiver side to form $K$ parallel interference-free channels.

\section{Alternative proof of Lemma~\ref{V1 and V2}}
\label{Alternative proof}
We provide an alternative proof of Lemma~\ref{V1 and V2}. Let $\text{diag}(\alpha_1, \alpha_2, \dots, \alpha_{n+1})= \mathbf{G}_{1,2} = \mathbf{H}_{1,1}^{-1}\mathbf{H}_{1,2}$ and $\text{diag}(\beta_1, \beta_2, \dots, \beta_{n+1}) = \mathbf{G}_{2,2} = \mathbf{H}_{2,1}^{-1}\mathbf{H}_{2,2}$. Therefore, all $\alpha$ and $\beta$ are independently drawn from a continuous distribution. We begin by looking at linear precoding matrix $\mathbf{V}_1$. In order to proof $\mathbf{V}_1$ is an invertible matrix, this time we show $\text{det}(\mathbf{V}_1) \neq 0$ almost surely. We express $\text{det}(\mathbf{V}_1)$ as
\begin{IEEEeqnarray}{rCl}
\text{det}(\mathbf{V}_1) & = &
\begin{vmatrix}
\alpha_1^n & \alpha_1^{n-1}\beta_1 & \dots  & \beta_1^n \\
\alpha_2^n & \alpha_2^{n-1}\beta_2 & \dots  & \beta_2^n \\
\vdots & \vdots & \ddots & \vdots \\
\alpha_{n+1}^n & \alpha_{n+1}^{n-1}\beta_{n+1} & \dots & \beta_{n+1}^n
\end{vmatrix}.
\end{IEEEeqnarray}

Let $A_{i,j}$ be the $(i,j)$ cofactor of $\mathbf{V}_1$. We expand $\text{det}(\mathbf{V}_1)$ along the first row such that
\begin{IEEEeqnarray}{rCl}
\label{determinant}
\text{det}(\mathbf{V}_1) 
& = &\alpha_1^n A_{1,1} + \alpha_1^{n-1}\beta_1 A_{1,2} + \dots + \beta_1^n A_{1,n+1}.
\end{IEEEeqnarray}
Notice that none of cofactor $A_{1,j}$ depends on $\alpha_1$ or $\beta_1$, where ${j \in \{1,2,\dots,n+1\}}$. If all values of cofactor $A_{1,j}$ and $\alpha_1$ are known, equation~(\ref{determinant}) is a polynomial equation in $\beta_1$. Now we proof $\text{det}(\mathbf{V}_1) \neq 0$ by contradiction. $\text{det}(\mathbf{V}_1) = 0$ implies $\beta_1$ is the root of the equation, or all $A_{1,j}$ and $\alpha_1$ are equal to zero. As $\beta_1$ is randomly drawn from a continuous distribution, it is not the root almost certainly. Then, all $A_{1,j}$ and $\alpha_1$ should equal to zero. 

We focus on the case that the $(1,1)$ cofactor $A_{1,1} = 0$. Cofactor $A_{1,1}$ can be simplified to the determinant of the submatrix formed by eliminating the first row and first column. We simplify $A_{1,1}$ as
\begin{align}
A_{1,1} =
\begin{vmatrix}
\alpha_2^{n-1}\beta_2 & \alpha_2^{n-2}\beta_2^2 & \dots  & \beta_2^n \\
\alpha_3^{n-1}\beta_3 & \alpha_3^{n-2}\beta_3^2 & \dots  & \beta_3^n \\
\vdots & \vdots& \ddots & \vdots \\
\alpha_{n+1}^{n-1}\beta_{n+1} & \alpha_{n+1}^{n-2}\beta_{n+1}^2 & \dots & \beta_{n+1}^n
\end{vmatrix}.
\end{align}
We expand the $A_{1,1}$ along the first row and keep repeating the above argument in a similar way. Finally, we can obtain $\beta_{n+1}^n = 0$. However, $\beta_{n+1}^n = 0$ is almost impossible because $\beta_{n+1}$ is drawn from a continuous distribution randomly. Hence, $\text{det}(\mathbf{V}_1) \neq 0$ and linear precoding matrix $\mathbf{V}_1$ is an ${(n+1) \times (n+1)}$ invertible matrix.

Owing to the matrix formed by the first $n$~rows of $\mathbf{V}_2$ has a similar structure to $\mathbf{V}_1$, $\mathbf{V}_2$ can be proved to have rank~$n$ in a similar way. This proof is omitted to avoid repetition.

\section{SNC scheme for General MIMO Cases}
\label{Degrees of Freedom for MIMO General Cases}
In this section, we look at the performance of our SNC scheme in general time-varying MIMO interference channels with limited receiver cooperation. There are $K$~transmitters and $L$~receivers, where $K$ and $L$ are not necessary identical. Transmitter $k$ has $M^{(T)}_k$ antennas and receiver~$l$ has $M^{(R)}_l$ antennas. 

Our SNC scheme for single antenna cases can be extended to general MIMO cases by considering each antenna of a node as a virtual node with an antenna. Our SNC scheme actives both $M = \min \left(\sum_{k=1}^K M^{(T)}_k, \sum_{l=1}^L M^{(R)}_l \right)$ virtual transmitters and virtual receivers. We consider each virtual receiver has an independent noiseless virtual cooperation link to the central processor. This can be achieved by allocating the resources of each cooperation link connected to its corresponding virtual receivers. We assume the capacity of each virtual cooperation link is just sufficient to forward the demodulated messages or the linear combination of the messages in the same finite field.

The general time-varying MIMO interference channel with limited receiver cooperation (as mentioned above) has total $M = \min \left(\sum_{k=1}^K M^{(T)}_k, \sum_{l=1}^L M^{(R)}_l \right)$ degrees of freedom. Our SNC scheme for single antenna cases described in Section~\ref{SNC in K-User Time-Varying Interference Channels with Limited Receiver Cooperation} can be extended to the general MIMO cases by considering there are $M$~virtual transmitters and $M$~virtual receivers. From the result in Section~\ref{SNC in K-User Time-Varying Interference Channels with Limited Receiver Cooperation}, this channel can achieve total $M$~degrees of freedom.




\ifCLASSOPTIONcaptionsoff
  \newpage
\fi

\end{document}